\documentclass[fleqn,12pt]{wlscirep}
\usepackage[T1]{fontenc}
\usepackage{times}
\usepackage{graphicx}
\usepackage{color}
\usepackage{ragged2e}
\usepackage{lineno}
\usepackage{xfrac}
\usepackage{siunitx}
\usepackage{lineno}
\usepackage{physics}

\DeclareSIUnit\au{a.u.}
\DeclareSIUnit\atomicunits{a.u.}
\DeclareSIUnit\torr{Torr}
\DeclareSIUnit\angstrom{\text{Å}}

\usepackage{booktabs}
\usepackage{amsmath,amssymb}

\title{1.5-Femtosecond Delay in Charge Transfer}

\author[1,$\dagger$]{Danylo T. Matselyukh}
\author[2,$\dagger$]{Florian Rott} 
\author[2,3,$\ast$]{Thomas Schnappinger}
\author[1,4]{Pengju Zhang}
\author[5]{Zheng Li}
\author[1]{Jeremy O. Richardson}
\author[2]{Regina de Vivie-Riedle}
\author[1,$\ast$]{Hans Jakob W\"orner}

\affil[1]{Department of Chemistry and Applied Biosciences, ETH Z\"urich, 8093 Z\"urich, Switzerland}
\affil[2]{Department of Chemistry, LMU Munich, 81377 Munich, Germany}
\affil[3]{Department of Physics, Stockholm University, AlbaNova University Center, SE-106 91 Stockholm, Sweden}
\affil[4]{Beijing National Laboratory for Condensed Matter Physics, IOP CAS, 100190 Beijing, People's Republic of China}
\affil[5]{School of Physics, Peking University, 100871 Beijing, People's Republic of China}

\affil[$\dagger$]{These authors contributed equally to this work}
\affil[$\ast$]{E-mail: thomas.schnappinger@fysik.su.se, hwoerner@ethz.ch}

\begin{abstract}
The transfer of population between two intersecting quantum states is the most fundamental dynamical event that governs a broad variety of processes in physics, chemistry, biology and material science. Whereas any two-state description implies that population leaving one state instantaneously appears in the other state, we show that coupling to additional states, present in all real-world systems, can cause a measurable delay in population transfer. Using attosecond spectroscopy supported by advanced quantum-chemical calculations, we measure a delay of 1.46$\pm$0.41~fs at a charge-transfer state crossing in CF$_3$I$^+$, where an electron hole moves from the fluorine atoms to iodine. Our measurements also fully resolve the other fundamental quantum-dynamical processes involved in the charge-transfer reaction: a vibrational rearrangement time of 9.38$\pm$0.21~fs (during which the vibrational wave packet travels to the state crossing) and a population-transfer time of 2.3--2.4~fs. Our experimental results and theoretical simulations show that delays in population transfer readily appear in otherwise-adiabatic reactions and are typically on the order of 1~fs for intersecting molecular valence states. These results have implications for many research areas, such as atomic and molecular physics, charge transfer or light harvesting. 

\end{abstract}

\begin{document}

\maketitle

\clearpage
The energetic crossing of quantum states is a general problem, critical for our understanding of most quantum systems' dynamics. In physics, relevant systems include nuclear spins in varying magnetic fields \cite{majorana1932atomi} (directly applicable to quantum information and computing), the dynamics of atoms and molecules in external electric or magnetic fields, particularly Stark-coupled Rydberg states \cite{rubbmark1981dynamical}, and the treatment of quantum decoherence and noise \cite{berns2006coherent}, while in chemistry these include charge transfer (CT) \cite{may11a,woerner17a}, the passage through avoided crossings (ACs) and conical intersections (CoIns) \cite{domcke2012role}. Generally, these crossings have been treated %
with a focus on determining the ultimate probability of diabatic/adiabatic population transfer \cite{ivakhnenko2023nonadiabatic} or the overall rate\cite{GRperspective}. Attosecond science opens an entirely new perspective on the dynamics of such crossings, allowing not only the final populations to be observed, but also the transient dynamical processes governing their evolutions. Applying such a time-resolved strategy to molecular charge transfer, this study identifies novel transient experimental observables that offer unique insights into the nature of state crossings. We show, both experimentally and theoretically, that even in the case of a direct two-state crossing of the type treated in the Landau--Zener--Stueckelberg--Majorana (LZSM) problem \cite{ivakhnenko2023nonadiabatic} or CT with Marcus theory~\cite{marcus93a}, additional intermediate states can influence the transient dynamics, leading to a unique observable: a delay in population transfer. Such novel observables can reveal the participation of previously neglected intermediate states in the quantum-state crossings of multi-level systems. These results have implications for the study of superconductivity, where the formation of Cooper pairs can be modulated by the superexchange interaction \cite{ruckenstein1987mean}, or revealing the mechanism (superexchange vs injection) driving CT along molecular bridges \cite{natali2014photoinduced}. %

For molecular valence-state crossings, we find these delays in population transfer to be on the order of a single femtosecond. We experimentally confirm and quantify such a delay by studying the sub-femtosecond dynamics of the $\tilde{\mathrm{B}}$ state of the trifluoromethyliodide cation (CF$_3$I$^+$), which undergoes considerable electron reorganization and charge transfer (CT) between photoionization and dissociation \cite{Yates1986-sw, aguirre2003ion}. These CT dynamics are probed by extreme ultraviolet (XUV) attosecond transient-absorption spectroscopy (ATAS), offering sub-femtosecond time resolution and indirect spatial sensitivity through its element specificity. ATAS is coupled with in-situ mass spectrometry, serving as an independent measure of the photoionization/dissociation, and ab-initio multiconfigurational second-order perturbation theory restricted active space (RASPT2) calculations, state-of-the-art calculations capable of accurately treating the spin-orbit coupling in both the valence- and core-ionized states of CF$_3$I$^+$. Our measurements reveal that strong-field ionization (SFI) with a carrier-envelope-phase (CEP) stable few-cycle pulse prepares a vibrational wave packet in the $\tilde{\mathrm{B}}$ state, which reaches a state crossing in 9.38$\pm$0.21~fs, the electronic-state population leaves (appears in) the $\tilde{\mathrm{B}}$ ($\tilde{\mathrm{E}}$) state with a timescale of 2.44$\pm$0.44~fs (2.34$\pm$0.36~fs).

In agreement with our theoretical model, we observe a non-instantaneous transfer of the population --- the appearance of the electron hole on the iodine atom (corresponding to the $\tilde{\mathrm{E}}$ state) is {\it delayed} by 1.46$\pm$0.41~fs with respect to its disappearance from the fluorine lone-pair orbitals (corresponding to the $\tilde{\mathrm{B}}$ state).
Apart from revealing this novel transient observable, our work represents the first time that the coupled electronic-nuclear processes of CT have been directly resolved and deconvolved. Previous measurements were able to estimate the overall timescale of state crossing/CT through exponential fits of experimental observables (e.g.\ yielding a 7-fs time scale for internal conversion in C$_2$H$_4^+$ \cite{zinchenko21a}) or by assuming kinetic rate laws (e.g.\ in core-hole clock experiments \cite{Schnadt2002}). Here, owing to an instrument-response function of $\sigma_\mathrm{Inst} = 1.00\pm 0.05$~fs (see Fig.~S2), we have been able to go beyond the traditional assumption of exponential kinetics.

Due to the ubiquity of state crossings in quantum processes, their treatment dates back to the advent of quantum mechanics with the LZSM transition \cite{ivakhnenko2023nonadiabatic}. Extensions of this treatment have increased the number of states that cross at a single point (``bow-tie'' crossing) \cite{ostrovsky1997exact}, or allow the treatment of successive crossings (``equal-slope'') \cite{demkov1968stationary} but studies of second-order interactions at state crossings (i.e.\ the coupling of two crossing states through a third) are scarce \cite{malla2022landau,PhilTransA}. Furthermore, to our knowledge, time delays in such processes have never been discussed. 
To investigate the effect that a third (intermediate) state has on the dynamics of a two-state crossing we introduce the following model Hamiltonian as a function of the reaction coordinate $x$: 
\begin{equation}
    \hat{H} =  \hat{K}(\hat{p}) + \hat{V}(x) = \left(\frac{\hat{p}^2}{2m}-\alpha_0 x\right) \mathbf{I}  + 
    \begin{bmatrix}
        0       & \gamma       & 0\\
        \gamma^* & c -\alpha x & \gamma \\
        0       & \gamma^*      & -2 \alpha x
    \end{bmatrix}
\end{equation}
where $\hat{K}$, $\hat{V}$, $\hat{p}$ are the kinetic-energy, potential-energy and momentum operators, respectively, $m$ is the mass, $\alpha_0$ is the gradient of the initial state, $\alpha$ is the gradient difference between the intermediate state and the crossing states, and $c$ and $\gamma$ are the energy offset and diabatic coupling (using atomic units throughout), respectively. The basis of diabatic electronic states therefore consists of the initial $\ket{0}$, intermediate $\ket{1}$, and final $\ket{2}$ states. Note that the direct coupling between the initial and final states is set to zero, but the presence of a small coupling does not fundamentally modify the results.

The resulting diabatic and adiabatic potential energy curves (PECs) are plotted in Fig.~1A and B for $\alpha_0 = 0.1$, $\alpha = 0.2$, $\gamma = 0.065$ and $c = 0.114$ (the mass is chosen to be the reduced mass of the CF$_3$ and I moieties). We investigate the temporal dynamics of the system by solving the time-dependent Schr\"odinger equation (TDSE) using the split-operator method \cite{leforestier1991comparison} (see Section~S1.2). We initialize the system (as a Gaussian centered at $x = -0.4$ with a probability width at half maximum of 0.192 and zero momentum) in the adiabatic ground state uphill of the crossing, which results in the wavepacket passing through the crossing, transferring population from $\ket{0}$ to $\ket{2}$. The transient population dynamics can be found in Fig.~1C. To investigate these systematically, they are fit with cumulative distribution functions (CDFs) of asymmetric generalized normal distributions (AGNDs)~\cite{Hosking1997}. These functions are chosen to capture the temporally-asymmetric nature of the rate of population change, which LZSM-like transitions are well known for. Section S1.1 discusses AGNDs and LZMS transitions in more detail.

\begin{figure}
    \centering
    \includegraphics[width=\textwidth]{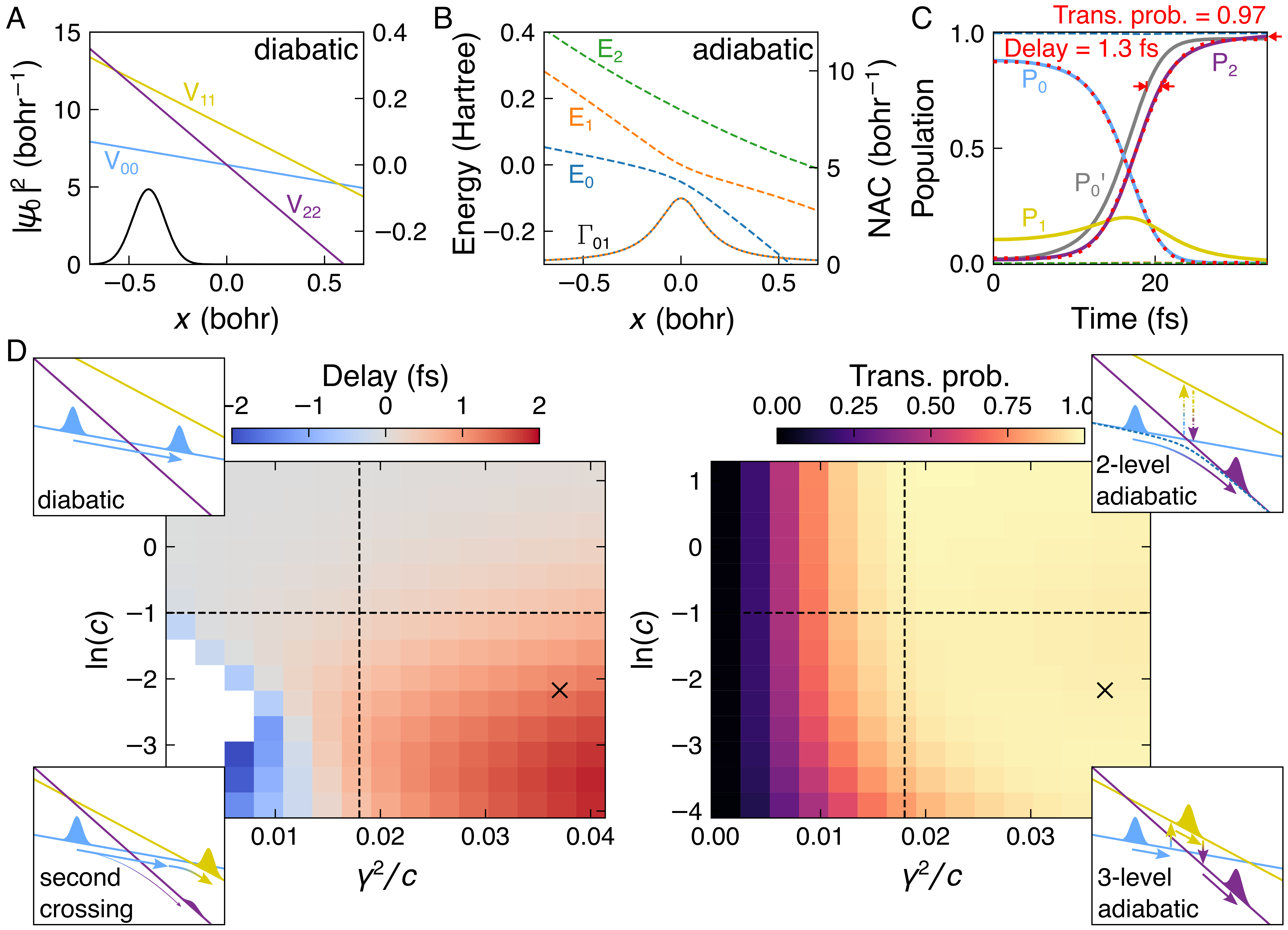}
    \caption{\textbf{Origin and magnitude of delays in population transfer}
    (\textbf{A}) A three-state crossing in the diabatic basis; the diagonal entries of $\hat{V}(x)$ are plotted alongside the initial wavefunction $\psi_0$ (black). (\textbf{B}) The AC between the bottom two states reveals itself in the adiabatic basis. The associated non-adiabatic coupling (NAC) $\Gamma_{01}$ is also shown (orange-blue dotted line). (\textbf{C}) The time-dependent populations resulting from a 1D TDSE propagation of $\psi_0$ according to $\hat{H}(x)$ are shown. Solid (dashed) lines show the diabatic (adiabatic) populations. $P_0$ (light-blue) and $P_2$ (purple) are fit with AGND CDFs (red dotted). To emphasize the 1.3~fs delay in population transfer, an amplitude-inverted and rescaled replica of the $P_0$ AGND fit is plotted (grey). The diabatic population transfer probability (taken to be the maximum $\ket{2}$ population over the 29.0~fs-long simulation) is also highlighted. (\textbf{D}) A parameter scan revealing the dependence of the delay and population transfer probability on the energy spacing $c$ and the Schrieffer--Wolf two-state coupling $\gamma_\mathrm{SW}/2 = \gamma^2/c$ (see Equation 2). Four distinct regions are identified and illustrated with inset schematic diagrams. A black cross marks the parameters of the simulation in (\textbf{A-C}).
    }
\end{figure}

For the system parameters presented in Fig.~1A--C, the population dynamics exhibit an important transient feature: a delay between the loss of population in $\ket{0}$ (light-blue) and gain in $\ket{2}$ (purple) due to it flowing through $\ket{1}$ (yellow). This delay is extracted from the difference of the position parameters $\mu$ of the fits. It is emphasized visually by inverting and rescaling the fit of the population in $\ket{0}$ such that it matches the asymptotic populations of the fit of the population in $\ket{2}$.
This `renormalization' of the fits shall be used throughout this work to visualize this novel population-transfer delay.

Figure~1D investigates how the population dynamics change as a function of $\gamma$ and $c$, revealing that the dynamics can be separated qualitatively into four regimes (separated by black dashed lines and illustrated with inset schematics). We find that when the offset $c$ is large, the system behaves as a two-state system with no apparent delay. This allows it to be treated with the Schrieffer--Wolff transform \cite{schrieffer1966relation} (see Section~S1.3) which removes the coupling to the intermediate state to lowest order and allows the Hamiltonian to be re-written as that of an effective two-state system:
\begin{equation}
    \hat{V} \rightarrow \hat{V}_\mathrm{SW} = -\alpha_0 x \mathbf{I} + 
    \begin{bmatrix}
        \frac{2 \gamma^2}{\alpha x - c}  & 0  & \frac{2 c \gamma^2}{\alpha^2 x^2 - c^2}\\
        0 & c -\alpha x - \frac{4\gamma^2 c}{\alpha^2x^2-c^2}& 0 \\
        \frac{2 c \gamma^2}{\alpha^2 x^2 - c^2}     & 0     & -2 \alpha x - \frac{2 \gamma^2}{\alpha x +c}.
    \end{bmatrix}
\end{equation}
In such cases, the dynamics are dominated by the effective coupling $\gamma_\mathrm{SW} = 2\frac{\gamma^2}{c}$ which determines whether the state crossing occurs diabatically (low transition probability, top-left quadrant) or adiabatically (high transition probability, top-right quadrant). In both cases, the population of $\ket{1}$ remains negligible, no delay in population transfer is present, i.e. the fits return the same location parameters $\mu_{\ket{0}} = \mu_{\ket{2}}$, and the same shape parameters $\kappa_{\ket{0}} = \kappa_{\ket{2}}$. In the case of CT, this regime corresponds to bridge-mediated superexchange \cite{natali2014photoinduced}.

When $c$ is lowered and becomes comparable to the wavepacket's kinetic energy at $x = 0$ ($K_{x=0}\approx \alpha_0 x_0$) it becomes possible to transfer population into $\ket{1}$ and the Schrieffer--Wolff transform fails. In the adiabatic regime (bottom-right quadrant and Fig.~1A), this leads to a unique observable --- a measurable delay in the population transfer $\mu_{\ket{0}} < \mu_{\ket{2}}$. Furthermore, the asymmetry of the population transfer, enforced to be equal by population conservation in the effective 2-state Schrieffer--Wolff limit, is now different for the initial and final states. In the case of bridge-mediated CT, this regime corresponds to injection \cite{natali2014photoinduced}. The bottom-left quadrant shows a negative delay due to a a second state crossing between $\ket{0}$ and $\ket{1}$ at $x>0$. Such dynamics fall under the category of ``triangular'' crossings \cite{kiselev2013su3} and are not relevant for the present discussion. 

To experimentally demonstrate the existence of population-transfer delays predicted by our model, we study the transient evolution of the $\tilde{\mathrm{B}}$ state of CF$_3$I$^+$. 
The electronic structure and energetic landscape of CF$_3$I$^+$ (presented in Fig.~2) is investigated with the help of RASPT2 ab-initio calculations (introduced below and in Section~S3): Fig.~2A presents the resulting diabatic PECs along with the dissociation limits of each state while Fig.~2B presents their leading electronic configuration, i.e. character. %
As expected, the calculations reveal the presence of a CT crossing between the $\tilde{\mathrm{B}}$ $^2\mathrm{A}_2$ state (hole in the lp$_{3}$ fluorine-lone-pair orbitals) and $\tilde{\mathrm{E}}$ $^2\mathrm{A}_1$ state (2-hole-1-particle configuration with a double hole in lp$_{1}$ and lp$_{2}$ and an electron in $\sigma^{*}_{5}$) very close to the Frank-Condon (FC) region, indicated by the SFI wavy arrow. The electronic characters, meanwhile, reveal that the studied CT requires the simultaneous rearrangement of two electrons as illustrated by dashed-dotted arrows in Fig.~2B. 

\begin{figure}
    \centering
    \includegraphics[width=\textwidth]{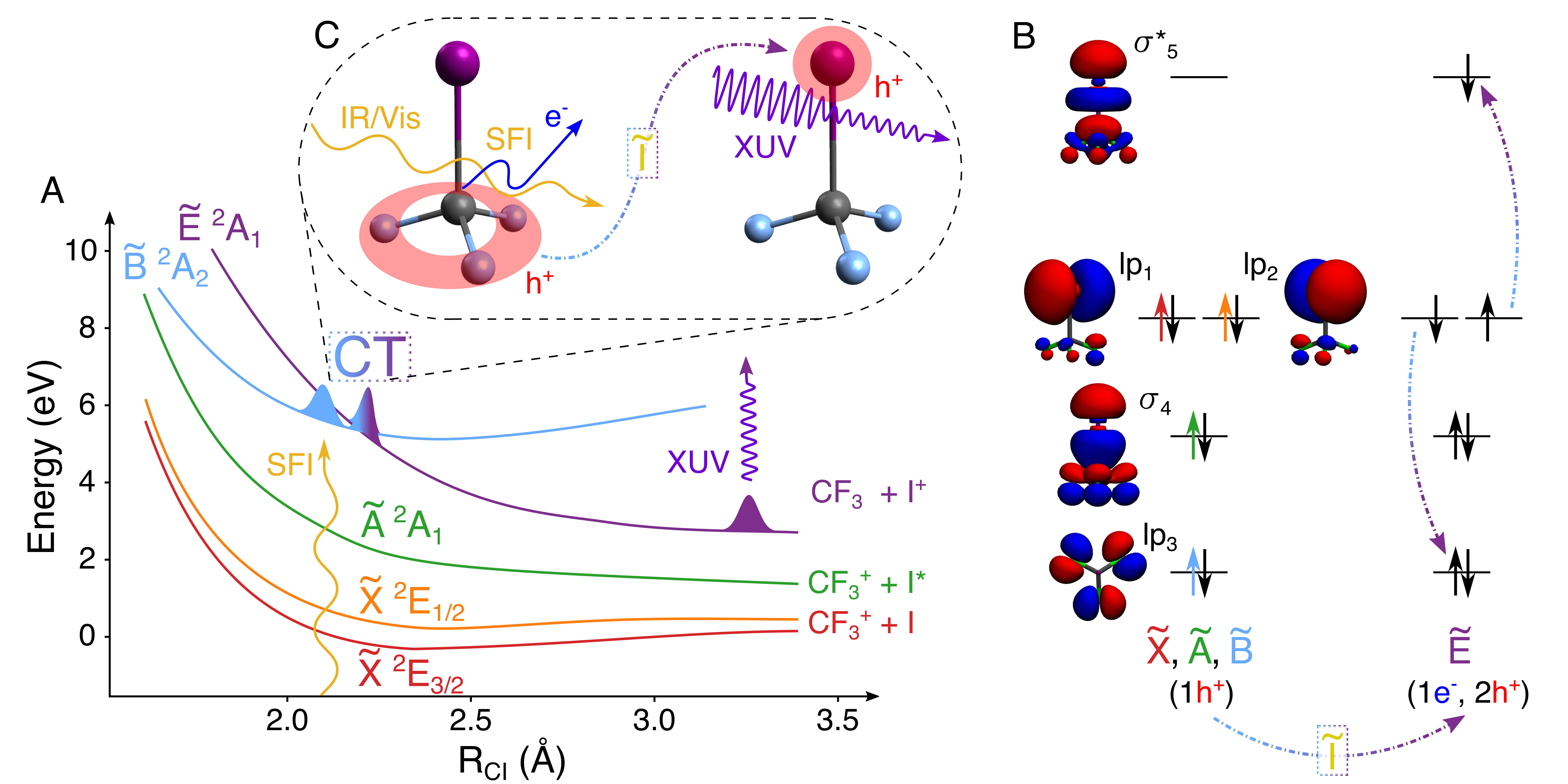}
    \caption{\textbf{CF$_3$I$^+$: a molecular system manifesting a charge-transfer crossing}
    (\textbf{A}) A schematic diabatic representation of the PECs of CF$_3$I$^+$ showing the initialization of the dynamics through SFI at the FC point and the nearby crossing of the $\tilde{\mathrm{B}}$ and $\tilde{\mathrm{E}}$ states.  (\textbf{B}) The leading character of the electronic states is shown with an orbital energy diagram (the respective ionized electrons of each state are shown as colored arrows). The passage between the $\tilde{\mathrm{B}}$ and $\tilde{\mathrm{E}}$ states requires the rearrangement of two electrons and therefore we postulate that it involves coupling to one or several intermediate states with $^2$E-symmetry (collectively labeled $\tilde{\mathrm{I}}$). (\textbf{C}) This rearrangement moves hole density from the fluorine atoms to the iodine which is observed through the site specificity of ATAS at the iodine N$_{4,5}$ edge.
    }
\end{figure}

Our attosecond experimental methods are illustrated in Fig.~3A--C and discussed in more detail in Section~S2. A CEP-stable 5.2-fs intense optical pump pulse is used to strong-field ionize CF$_3$I, while a time-delayed isolated attosecond pulse probes the ensuing dynamics through iodine-4d core-to-valence transitions (between 45 and 55~eV). Fluctuations in the attosecond probe pulse's spectrum were compensated using singular-value-decomposition-based filtering~\cite{matselyukh22a, matselyukh2023attosecond} (see Section S2.3). The temporal confinement of SFI~\cite{wirth11a,pabst16a} results in a measured experimental cross-correlation of $\sigma_\mathrm{Inst} = 1.00\pm0.05$~fs (Fig.~S2). SFI simultaneously prepares multiple electronic states of the cation, initiating different vibrational dynamics in each. Note that the quasi-static (i.e. electronically-adiabatic) nature of SFI in small molecules \cite{ammosov1986tunnel} is the reason why we initialize all of our three-state simulations in the adiabatic ground state. 

The dynamics in different electronic states of CF$_3$I$^+$ are deconvolved by coupling our ATAS beamline~\cite{huppert15a,matselyukh22a} with an in-situ time-of-flight mass spectrometer (TOF-MS, Fig.~3C). This device identifies the ionic fragments (Fig.~3D) generated by the pump pulse independently of experimental and simulated absorption spectra (see Section~S2.2). By tuning the pump-pulse intensity through an in-vacuum motorized iris, we acquire intensity-resolved ion yields and transient-absorption spectra (Fig.~3E,F), which allow us to determine the optimal peak intensity for investigating the $\tilde{\mathrm{B}}$ state dynamics with ATAS. An intensity of 300~TW$\cdot$cm$^{-2}$ gives a large fraction of I$^+$ (and therefore the initial $\tilde{\mathrm{B}}$ state population) while minimizing the abundance of fragments like CF$_2^+$, which originate from higher-lying excited states.

\begin{figure}
    \centering
    \includegraphics[width=\textwidth]{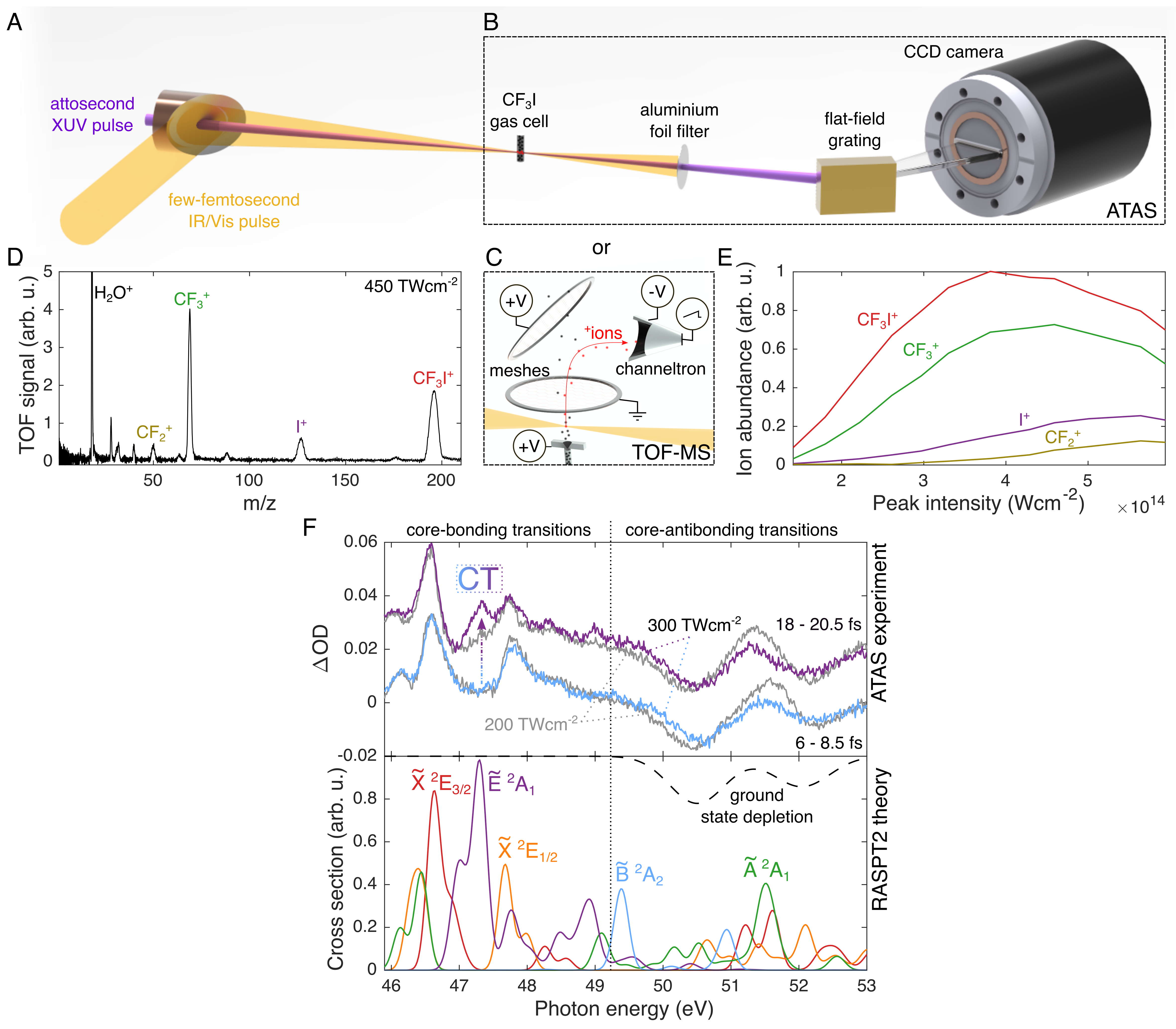}
    \caption{\textbf{Experimental methodology and overview of results.}
    (\textbf{A}) An attosecond beamline~\cite{huppert15a} delivers an isolated attosecond XUV pulse and a few-femtosecond near-infrared/visible (IR/Vis) pulse with a tunable delay. These are used to perform (\textbf{B}) ATAS~\cite{matselyukh22a}, or (\textbf{C}) TOF-MS measurements. (\textbf{D}) Mass spectrum measured at a peak intensity of 450~TW$\cdot$cm$^{-2}$. The relative intensity-dependent abundancies of the major fragments are shown in (\textbf{E}).
    (\textbf{F}) Two transient spectra from two ATAS experiments at different pump intensities (top) are compared with energy-shifted RASPT2 ab-initio spectra (see Section~S3.4) computed at a C--I bond length of 2.34~\AA~(bottom). An intensity- and delay-dependent feature is observed at 47.3~eV corresponding to the CT $\tilde{\mathrm{E}}$ state. 
    }
\end{figure}

The experimental results are interpreted with the help of state-of-the-art restricted-active-space self-consistent-field (RASSCF) methodology~\cite{Malmqvist1990-fk,Roos2009-ku}, used to calculate the valence-excited states of CF$_3$I$^+$ as well as their core-valence absorption spectra, while also capturing the effects of spin-orbit coupling. The general protocol for calculating these states and transitions up to second-order perturbation theory (RASPT2) is discussed in Ref.~\cite{Rott2021-vb} and Section~S3. To treat CF$_3$I$^+$, the restricted active space RAS($34,1,1;5,13,3$) illustrated in Figure~S8 was utilized. Here the orbital space is divided into three sub-spaces, RAS1, RAS2 and RAS3. As the ATAS probed the $N_{4,5}$ edge of iodine, its five $4d$ orbitals were included in the RAS1, and a maximum of one electron was allowed to be excited out of this sub-space. The RAS2 was comprised of the carbon-iodine bond ($\sigma_{4}$, $\sigma^{*}_{5}$), both iodine lone-pair orbitals lp$_{1}$ and lp$_{2}$ as well as the six fluorine lone-pair orbitals lp$_{3}$, lp$_{4}$, lp$_{5}$, lp$_{6}$, lp$_{7}$ and lp$_{8}$. Additionally it included the three carbon-fluorine bonds $\sigma_{1}$, $\sigma_{2}$, and $\sigma_{3}$, but to reduce computational costs, their corresponding virtual orbitals ($\sigma^{*}_{6}$, $\sigma^{*}_{7}$ and $\sigma^{*}_{8}$) were included in the RAS3 sub-space and only a single excitation into them was allowed, while for the orbitals in the RAS2 all possible configurations were taken into account. This partitioning of the orbital space makes the RAS formalism conceptually a very intuitive and elegant way to describe the core excitation process. Further details on the calculation of the PECs are provided in Section~S3.

Figure~3F compares the transient-absorption spectra measured at two pump-probe delays and pump intensities with the RASPT2 calculations, demonstrating quantitative agreement. At the lower (200~TW$\cdot$cm$^{-2}$) pump intensity, where only the spin-orbit-split $\tilde{\mathrm{X}}$ $^2$E ionic ground state is excited (grey line), the core-bonding region (<49~eV) exhibits four absorption features; %
two weaker at 46.1 and 48.4~eV and two stronger at 46.6 and 47.8~eV.
Their respective splittings correspond to the 1.7~eV (0.6~eV) intervals of the 4d (5p) levels of iodine. The absence of observable signatures of the $\tilde{\mathrm{A}}$ state is discussed in Section~S4. 

When the pump intensity is increased to 300~TW$\cdot$cm$^{-2}$, a new delayed absorption feature appears at 47.3~eV  (highlighted with `CT' in Fig.~3F) along with weaker features between 48 and 49.5~eV. These match the RASPT2 spectrum of the $\tilde{\mathrm{E}}$ state, as well as previous measurements of the N$_{4,5}$ absorption of I$^+$~\cite{OSullivan1996} (to which the $\tilde{\mathrm{E}}$ state dissociates). The agreement of experiment, literature, and calculations allows for the unambiguous assignment of the absorption peak at 47.3~eV to the $\tilde{\mathrm{E}}$ state.

Since the hole density in the $\tilde{\mathrm{B}}$ state is localized on the fluorine atoms with very little spatial overlap with the iodine 4d orbitals, its spectrum contains minimal core-to-hole-type transitions, which dominate the N$_{4,5}$ spectra of the $\tilde{\mathrm{X}}$, $\tilde{\mathrm{A}}$ and $\tilde{\mathrm{E}}$ states. Instead, similar to the CF$_3$I neutral ground state (the spectrum of which can be found in Fig.~S6), the $\tilde{\mathrm{B}}$-state spectral features are weaker and limited to the core-antibonding and core-Rydberg transitions which are found at spectral energies $>$49.5~eV (as shown in Fig.~3F).

To identify these $\tilde{\mathrm{B}}$-state absorption features and investigate their temporal evolution we turn to Fig.~4 which compares the high-pump-intensity ATAS results (A) with the calculated absorption spectra of the first, second and fourth adiabatic states (B). The calculated spectra are also plotted separately using a common color scale in Fig.~S12. Time zero ($\Delta t = 0$~fs) is set to the center of rise of the 46.7-eV $\tilde{\mathrm{X}}\,^2\mathrm{E}_{3/2}$-state absorbance, determined from an error function fit $0.0\pm 0.3$~fs (see bottom of Fig.~4A). The previously discussed onset of the $\tilde{\mathrm{E}}$-state absorption around 10~fs is again evident. This is also reproduced in the RASPT2 calculations where the fourth adiabat only exhibits $\tilde{\mathrm{E}}$-state absorption after the CT crossing at $R_\mathrm{CI} = 2.21$\,\AA. 

Figure~4A also reveals the presence of a weak, broad and short-lived absorption feature between 53.5 and 55~eV. Although slightly shifted compared to the calculated $\tilde{\mathrm{B}}$-state spectra, this feature disappears just as the absorption at 47.3~eV rises, indicating that population is being transferred between the two states. The bottom panel of Fig.~4A shows this anti-correlation more clearly, plotting the integrated transient absorption as a function of time for the regions 46.36--47.09~eV (I), 47.09--47.53~eV (II), 47.53--48.12~eV (III) and 53.50--54.30~eV (IV). This provides sufficient evidence to assign feature IV to the $\tilde{\mathrm{B}}$ state, and together with the nuclear-coordinate-invariance of the state-averaged RASPT2 transition dipole (demonstrated in Section~S3.6), allows regions II and IV to be used as measures of the transient diabatic populations of the $\tilde{\mathrm{B}}$ and $\tilde{\mathrm{E}}$ states. The population transfer delay can now be experimentally investigated.

\begin{figure}
    \centering
    \includegraphics[width=\textwidth]{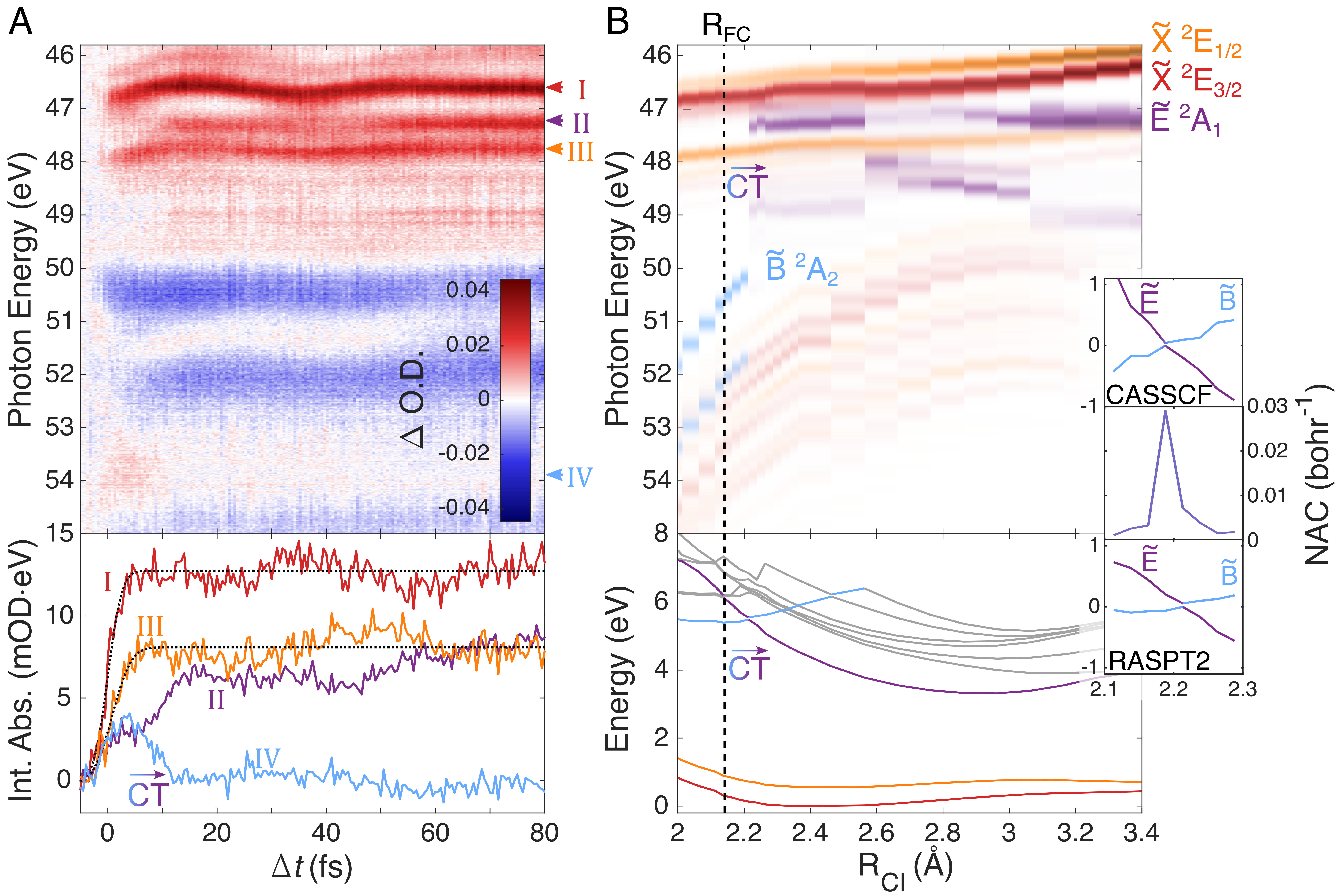}
    \caption{\textbf{Detailed analysis of ATAS results}
    (\textbf{A}) The top panel shows the ATAS results of the high-intensity measurement (300~TW$\cdot$cm$^{-2}$). Colored arrowheads on the right identify the spectral features whose integrated transient absorption is investigated in the bottom panel (see the main text for the spectral ranges used). %
    (\textbf{B}) The top panel shows the calculated energy-shifted RASPT2 $R_\mathrm{CI}$-resolved absorption spectra (see Section~S3.4) of the relevant cationic states. The spectra of the different states are differentiated by color and the FC point ($R_\mathrm{FC} = 2.14$\,\AA) is marked with a vertical dashed line. 
    In the bottom panel, the adiabatic PECs of CF$_3$I$^+$ are shown, color-coded according to their leading electronic character. The $\tilde{\mathrm{B}}$ / $\tilde{\mathrm{E}}$ crossing at $R_\mathrm{CI}$ = 2.21\,\AA~is evident. The inset (right) shows this CT crossing in more detail for the CASSCF and RASPT2 calculations.
    }
\end{figure}

We extract the delay in the same manner as for the results of the simulations; by fitting the CDF of the AGND to the transient absorption (details in Section~S2.4). The resulting location parameters $\mu_{\tilde{\mathrm{B}}} = 8.65\pm 0.34$~fs and  $\mu_{\tilde{\mathrm{E}}} = 10.11\pm 0.23$~fs reveal a significant population-transfer delay: $\tau = \mu_{\tilde{\mathrm{E}}} - \mu_{\tilde{\mathrm{B}}} = 1.46 \pm 0.41$~fs. 
A two-state diabatic model is unable to replicate such dynamics; meanwhile, our three-state simulations confidently reproduce the delay (see Fig.~5B and C). As in Fig.~1, the dynamics are almost entirely adiabatic, exhibiting a population transfer delay of 1.48~fs. The initial vibrational wavepacket was chosen to match the ground state of CF$_3$I (with the same width as the simulations in Fig.~1 but initially centered at $-0.12$~\AA). The parameters of the Hamiltonian were $\alpha_0 = 0.25$, $\alpha = 0.2$, $\gamma = 0.045$ and $c = 0.04$. 

It is important to note that vibronic coupling between $\tilde{\mathrm{B}}$, $\tilde{\mathrm{I}}$ and $\tilde{\mathrm{E}}$ is symmetry-forbidden for geometries in the C$_{3v}$ point group (as is the case for the RASPT2 calculations) because the direct product of pairs of these states and the C-I stretching mode are not totally symmetric. The CT crossing shown in the PECs is thus a true CoIn. An adiabatic reaction can only proceed by passing around it through symmetry breaking, which naturally occurs due to the finite extension of the vibrational wavepacket along non-totally-symmetric modes. Our three-state model actually describes the CT-reaction at such distorted geometries. A detailed symmetry analysis of this problem is provided in Section~S5.

The experimental AGND fits allow us to also determine the vibrational rearrangement time of the CT reaction from the mean of the location parameters $(\mu_{\tilde{\mathrm{B}}} + \mu_{\tilde{\mathrm{E}}})/2 = 9.38 \pm 0.21$~fs. Meanwhile, the scale parameters $\sigma_{\tilde{\mathrm{B}}} = 2.44\pm 0.44$~fs and $\sigma_{\tilde{\mathrm{E}}} = 2.34\pm 0.36$~fs represent the times taken for the initial diabatic state to lose its population and for the final state to gain it, respectively. To date, such timescales could not be deconvolved from the vibrational dynamics (rearrangement time) and only be inferred indirectly, e.g. through core-hole clock spectroscopy \cite{Schnadt2002}.

\begin{figure}
    \centering
    \includegraphics[width=\textwidth]{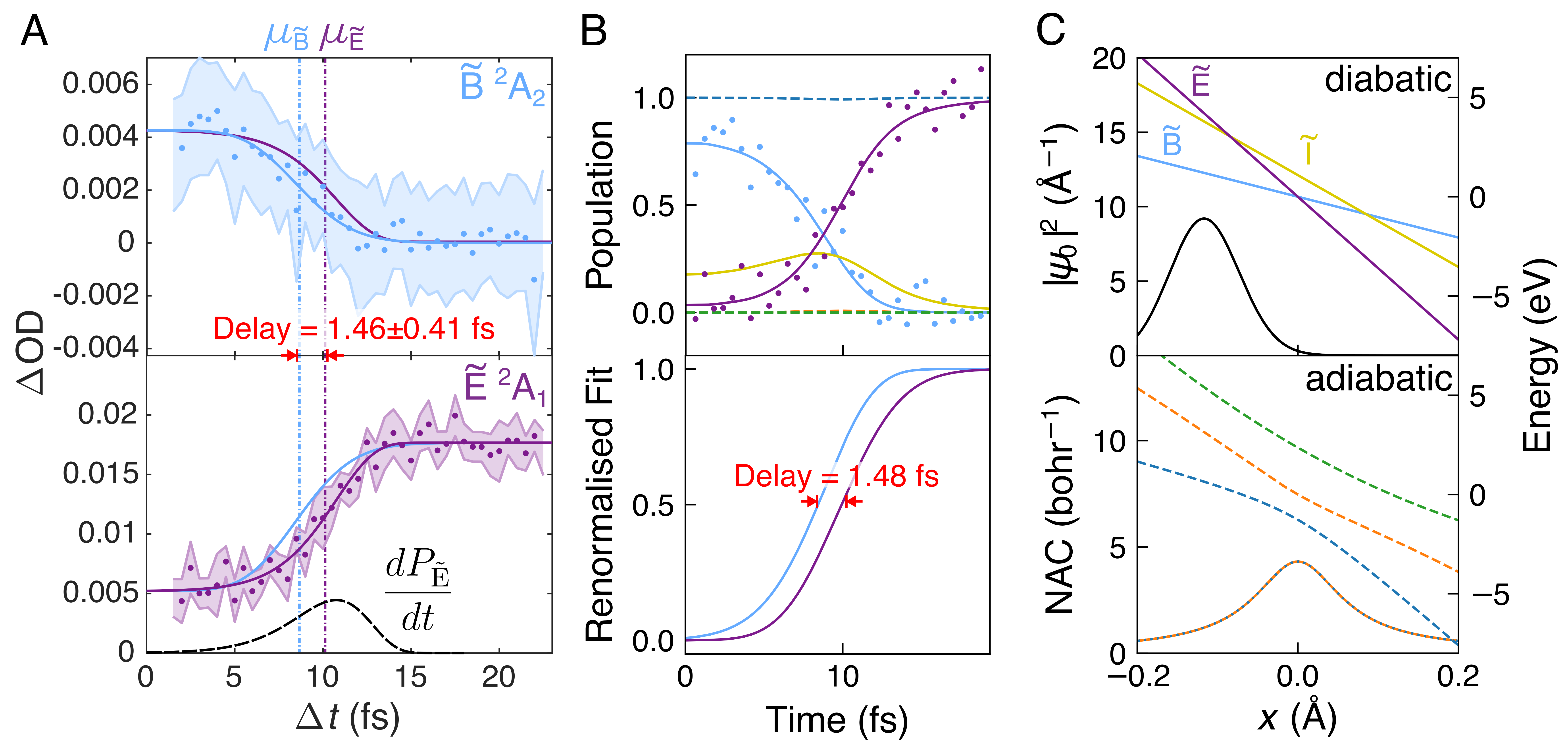}
    \caption{\textbf{Determination and theoretical reproduction of the charge-transfer delay.}
    (\textbf{A}) Transient absorption of the $\tilde{\mathrm{B}}$ and $\tilde{\mathrm{E}}$ states (dots, shaded error ranges represent the standard deviations) fit with the AGND CDF (solid lines). %
    The location parameters, $\mu_i$, are shown as dashed-dotted vertical lines. 
    Lines of opposite color to the dots show the inverted and rescaled fit of the other state to emphasize the delay. The AGND probability density function of $\tilde{\mathrm{E}}$, corresponding to the rate of population transfer is plotted below as a dashed black line. %
    (\textbf{B}) The top panel shows how the rescaled experimentally observed population dynamics (dots) can be reproduced with the three-state crossing model introduced in Fig.~1 (the diabatic (solid) and adiabatic (dashed) transient populations are shown). The delay in the population transfer is emphasized in the bottom panel which shows the renormalized AGND fits of the simulated population dynamics. The effective one-dimensional PECs, initial probability distribution and NAC used for these calculations are shown in (\textbf{C}).
    }
\end{figure}

Starting with a general theoretical model we have shown the existence and elucidated the origin of a novel transient observable in state crossing dynamics: a delay in population transfer caused by the coupling of two crossing states through a third. Using attosecond spectroscopy supported by advanced quantum-chemistry calculations, we have also demonstrated its existence experimentally. Investigations of the theoretical model's parameters have revealed that such delays can be expected to be on the order of a single femtosecond for intersecting valence states of molecules, opening an exciting frontier for attochemistry.

The site specificity and sub-femtosecond temporal resolution achieved in this work have moreover enabled the direct real-time observation of the primary quantum-mechanical processes underlying molecular charge transfer. Our measurements have determined the vibrational rearrangement time taken by the nuclear wave packet to reach the CT crossing (9.38$\pm$0.21~fs), the population-transfer times (2.3--2.4~fs) and a delay (1.46$\pm$0.41~fs) in population transfer. In contrast to charge migration, the time scale of which is simply determined by the energetic intervals of the populated quantum states, the delay discovered in this work encodes diabatic couplings to additional states beyond the two intersecting states. These couplings are neither accessible to frequency-domain spectroscopy, nor straightforward to calculate. As such, these delays reflect non-trivial molecular properties and affect one of the most fundamental quantum processes in matter.

Looking forward, our work defines a set of methodologies for unraveling the primary quantum-mechanical processes that underlie all population-transfer dynamics, in particular long-range charge transfer \cite{schuster00a}, which is known to play a fundamental role in photovoltaics \cite{gelinas14a}, photosynthesis \cite{lee07a,collini10a}, and photocatalysis \cite{park16a}. The sensitivity of XUV/X-ray spectroscopy to electronic coherences \cite{goulielmakis10a,matselyukh22a} and its applicability to complex liquid-phase systems \cite{smith20a,yin23a} make the present methodology relevant for probing and exploiting coherences in complex molecular systems featuring scientifically and technologically relevant charge-transfer dynamics, such as those underlying photosynthesis, photovoltaics or molecular optoelectronics.

\section*{Acknowledgements}

We thank M. Belozertsev for his early contributions to the theoretical research and A. Schneider, M. Seiler and M. Urban for their technical support. D.M., P. Z. and H.J.W. gratefully acknowledge funding from an European Research Council Consolidator grant (project no. 772797-ATTOLIQ), as well as projects no. 200021\_172946 and the NCCR-MUST, funding instruments of the Swiss National Science Foundation. F.R., T.S. and R.d.V. acknowledge funding from the DFG Normalverfahren and the computational and data resources provided by the Leibniz Supercomputing Centre (www.lrz.de). Z.L. acknowledges funding from National Natural Science Foundation of China (project no. 12174009, 12234002).

\paragraph*{Author Contributions:} H.J.W and D.M. conceived the experimental study. D.M. and P.Z. performed the experiments and D.M. analyzed the results. D.M. and J.O.R. performed the three-state model simulations. F.R. and T.S. performed, analyzed and post-processed the RASPT2 calculations. H.J.W. supervised the experimental work. T.S. and R.d.V. supervised and coordinated the quantum-chemistry part of this work. Z.L. and D.M. performed preliminary quantum-chemical calculations. D.M., F.R., T.S., J.O.R., H.J.W and R.d.V. discussed the results, finalized the interpretations and wrote the paper. All authors reviewed and/or edited the paper.

\paragraph*{Competing Interests:} The authors declare that they have no competing interests.

\paragraph*{Data and Materials Availability:} The data that support the findings of this study will be uploaded to the public ETH e-collection repository upon acceptance of this manuscript.

\section*{Supplemental Materials}
Theoretical, Experimental and Computational Details\\
Additional discussions\\
Figures S1 to S12\\
Tables S1 to S4\\

\end{document}

\maketitle
\tableofcontents
\newpage

\section{Theoretical Details}

\subsection{Origin and treatment of the asymmetric population transfer}

The most basic crossing of two states in time is well understood within the Landau--Zener--Stueckelberg--Majorana (LZSM) treatment already introduced in the main text. Thanks to exact solutions of the problem as well as numerical and approximate treatments \cite{Vutha2010}, it is know that the typical time evolution of the diabatic populations follows two phases; an initial jump in population which starts slow, but rapidly accelerates and reaches the asymptotic population. After this, a period of relaxation can follow, during which oscillations in population can sometimes be seen. One qualitative way of interpreting this evolution is to consider that the rate of population transfer depends on the population in the initial state; as the population of the initial state is rapidly depleted, so does the rate of population transfer. 

As a result of the gradual start and rapid cessation of the dynamics, the rate of diabatic population transfer $\frac{dP_i}{dt}$ exhibits a clear asymmetry. Therefore, to achieve a robust fit of our simulated (and experimental) population dynamics the function used to fit them must possess an adjustable asymmetry. One of the most general functions of this type whose parameters are entirely uncoupled is the asymmetric generalized normal distribution (AGND)~\cite{Hosking1997}. Its cumulative distribution function (CDF) takes the form 
\begin{equation}
A(t)= \frac{a}{2}\left[ 1 + \mathrm{erf}\left(\frac{y(t)}{\sqrt{2}} \right) \right] + c, 
\end{equation}

where

\begin{equation*}
    y(t) = 
    \begin{cases}
        -\frac{1}{\kappa} \log \left(1 - \frac{\kappa(t-\mu)}{\sigma} \right) & \text{if } \kappa \neq 0\\
        \frac{t-\mu}{\sigma} & \text{if } \kappa = 0
    \end{cases}
    ,
\end{equation*}
$a$ is the amplitude, $c$ is the offset, $\mathrm{erf}()$ is the error function, $\mu$ is the location parameter, $\sigma$ is the scale parameter and $\kappa$ is the shape parameter.

The asymmetry that the shape parameter $\kappa$ describes is generally known as `skew'. Functions with a long tail in the negative direction (as is the case for the rate of diabatic population transfer) are said to be `negatively skewed' and exhibit $\kappa>0$.

\subsection{Simulating the 3-state model}

To solve the time-dependent Schrödinger equation (TDSE) for the 3-state model system presented in our work we use the split-operator method. Convergence of the simulation results was verified with respect to grid size and time step. A grid spacing of 0.002 a.u. and a time step of 5 a.u. were found to be suitable. To ensure that the populations were conserved, no absorbing boundary conditions were employed and the grid was made large enough that the wave packet did not reach the edge within the simulation time.

\begin{figure}
    \centering
    \includegraphics[width=0.7\textwidth]{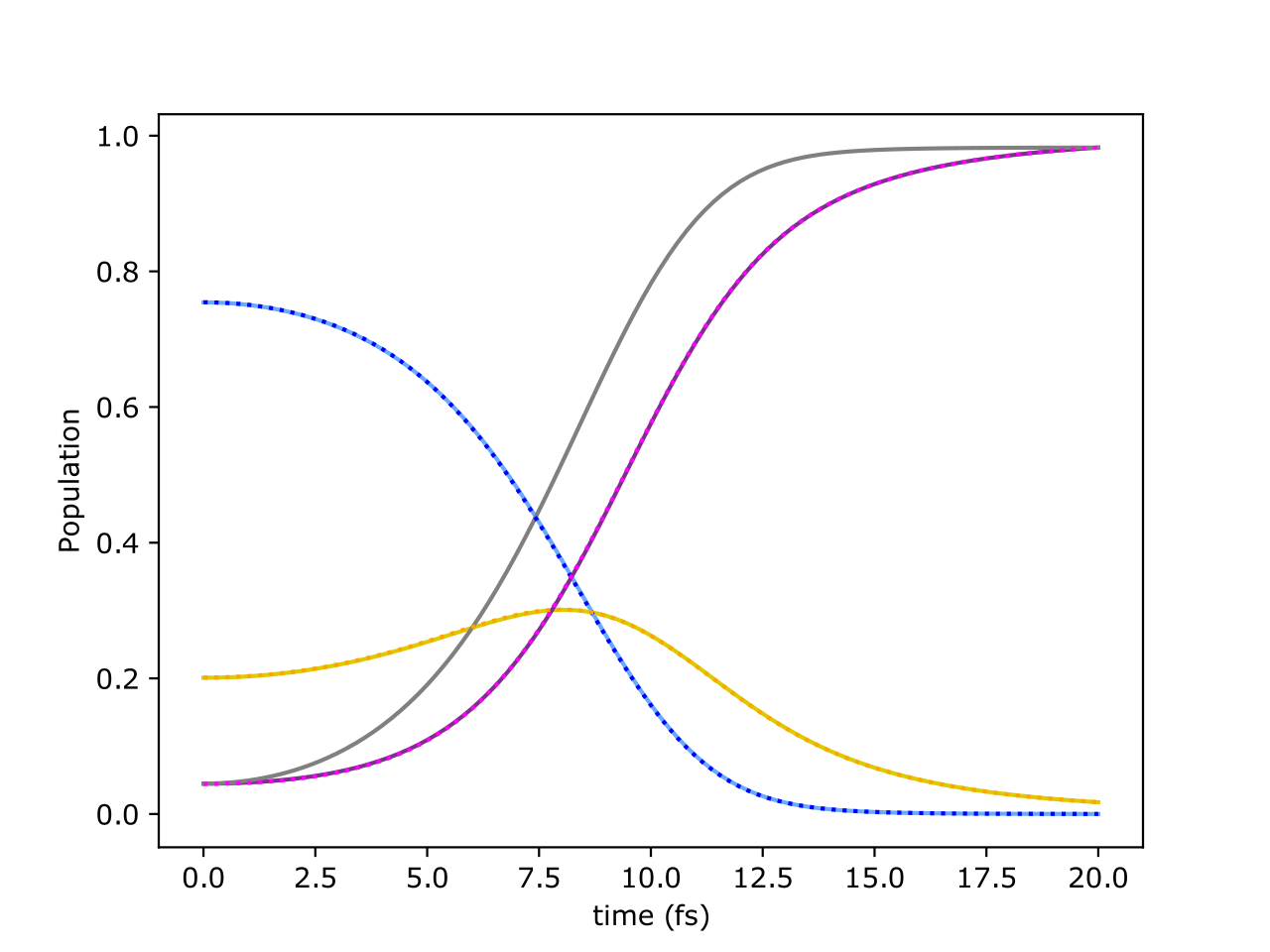}
    \caption{Comparison between the quantum and classical treatment of three-state model in the adiabatic limit. The parameters used for the simulations are the same as in Fig.~5. The line colors represent the three diabatic states; initial (blue), intermediate (yellow) and final (purple). The classical results are plotted as solid lines and the quantum results as dotted lines. The solid gray line shows the rescaled initial state population from the classical calculation. A population transfer delay is still observed. }
    \label{fig:classical}
\end{figure}

It is worth noting that for simulations occurring deep in the adiabatic regime, i.e. in the right two quadrants of Fig.~1D, the results were practically unchanged when propagating the wavepacket on the ground-state Born--Oppenheimer potential. We also found almost perfect agreement when using an ensemble of classical trajectories initialized from a Wigner transform of $\psi_0$ and evolved on the ground-state Born--Oppenheimer potential (see Fig.~\ref{fig:classical}). At each timestep of the classical simulations, the diabatic populations were calculated using the square of the elements of the ground-state eigenvector of the diabatic Hamiltonian.
These results suggests a direction that could be taken in future work to simulate a multi-state crossing and the associated delay from first principles.

\subsection{The Schrieffer--Wolf transform}

The Schrieffer--Wolf transform is able to decouple one sub-system from interactions with an energetically-distant sub-system. When applied to our model Hamiltonian, it can be used to find the effective `direct' coupling between the initial $\ket{0}$ and final state $\ket{2}$. 

To do this, the Hamiltonian $\hat{H}$ is first decomposed into a diagonal part $\hat{H}_0$ and the coupling 
\begin{equation*}
    \hat{V'} = \begin{bmatrix}
        0       & \gamma       & 0\\
        \gamma & 0 & \gamma \\
        0       & \gamma      & 0
    \end{bmatrix}.
\end{equation*}
This coupling can be removed from the system Hamiltonian to first order in $\hat{V'}$ by finding a generator $S$ that satisfies:
\begin{equation}
    [\hat{H}_0,S] = \hat{V}',
\end{equation}
such that 
\begin{align*}
    \hat{H}' = e^{S}\hat{H}e^{-S} &= \hat{H} + [S,\hat{H}] + ...\\
    &= \hat{H}_0 + \hat{V}' + [S, \hat{H}_0] + [S, \hat{V}'] + ...\\
    &= \hat{H}_0 + [S, \hat{V}'] + ...
\end{align*}

Solving (S2) for $S$, we find
\begin{equation}
    S = \begin{bmatrix}
        0       & \frac{\gamma}{\alpha x - c}       & 0\\
        \frac{-\gamma}{\alpha x - c}  & 0 & \frac{\gamma}{\alpha x + c}  \\
        0       & \frac{-\gamma}{\alpha x + c}      & 0
    \end{bmatrix},
\end{equation}
allowing the Schrieffer--Wolf Hamiltonian to first order in $\hat{V}'$ to be found (see Equation 2 of the main text).

\section{Experimental Details}

\subsection{Experimental methods}

The optical setup of the experiment begins with a FEMTOPOWER V CEP laser system, which delivers 1.5~mJ, 25~fs laser pulses centered at 790~nm at a repetition rate of 1~kHz. These pulses are spectrally broadened in a 1~m long hollow-core fibre filled with neon, producing an octave spanning spectrum. Using eight bounces off PC-70 Ultrafast Innovations mirrors and a pair of fused silica wedges positioned at Brewster's angle to the beam, the supercontinuum is compressed to 5.2~fs. The duration of the pulse is characterized using a home-built second harmonic generation D-scan device. 

The few-cycle pulse is used for both, generating the extreme-ultraviolet (XUV) isolated attosecond probe-pulse, and directly as a few-cycle pump-pulse. This is achieved by recycling the residual few-cycle driving field after the isolated attosecond pulse generation. The XUV attosecond pulse is generated through high-harmonic generation (HHG) in a differentially pumped finite gas cell filled with argon, after which the visible and XUV light is split using a drilled parabolic mirror. The reflected and collimated optical light then passes through a under-vaccuum, actively stabilized delay line, becoming the pump pulse. The 26~as stability of the delay line is achieved using two piezo-driving PID control loops for which the error signal is produced by a He:Ne based interferometer. The probe pulse is spectrally filtered using few-hundred-nanometer-thick aluminum foil which reflects the visible light while transmitting the XUV. A toroidal mirror is used to refocus the diverging XUV probe pulse into the transient absorption spectroscopy (TAS) target in a 2f-2f geometry, passing through another drilled parabolic mirror in the process. This parabolic mirror collinearly recombines the pump and probe-pulses, while also focusing the pump pulse into the TAS target. The resulting 

\begin{figure}
    \centering
    \includegraphics{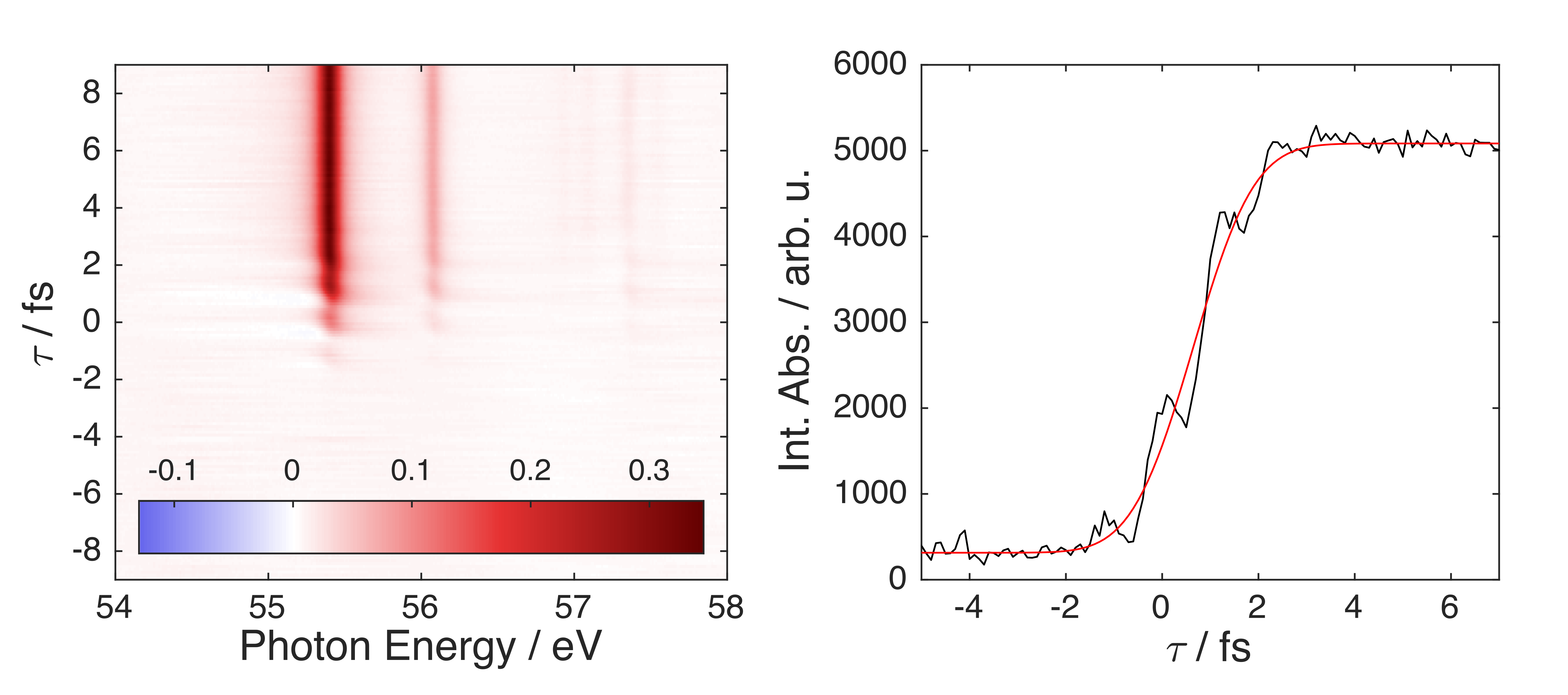}
    \caption{Determination of the instrument response function. Left: The results of ATAS on xenon atoms under experimental conditions very similar to the present work are shown in a pseudocolor plot; the measured change in optical density ($\Delta$OD) is plotted as a function of photon energy and delay $\tau$. The appearance of three absorption lines due to SFI of xenon are evident.%
    Right: The integrated $\Delta$OD over the strongest absorption line at 55.4~eV ($^2$P$_{3/2}\rightarrow^2$D$_{5/2}$ of Xe$^+$, 5p$^{-1}$) is fit with a error function determining the experimental cross-correlation standard deviation to be 1.00$\pm$0.05~fs. This figure is reproduced with permission from \cite{matselyukh2023attosecond}.}
    \label{fig:enter-label}
\end{figure}

Two different TAS targets were used in this work, but their performance was similar. The first design was a 3~mm tube fed with an Even-Lavie pulsed valve, the second a 1~cm cell continuously backed with the target gas.

Upon passing the transient absorption target, the pump and probe are separated with another aluminum-foil filter. The transmitted XUV was spectrally dispersed using a Hitachi 001-0660*3 aberration-corrected concave grating (positioned with the TAS target at its imaging focus) and detected using a Princeton Instruments PIXIS-XO:2KB backlit-CCD, providing a resolving power over 1000. 

Before running experiments on CF$_3$I, the XUV spectrometer was spectrally calibrated by measuring the Fano resonances in the photoionization continuum of neon which appear between 45 and 52~eV. Once calibrated, the transient-absorption target was supplied with 99\% purity CF$_3$I purchased from Apollo Scientific Limited and TAS was carried out by varying the delay between the pump and probe pulses using the delay line. The absolute delay was also simultaneously tracked with the help of a white-light interferometer. The pump intensity was varied with the use of a motorized iris placed in the delay line. 

In addition to their XUV absorption spectrum, the ions generated by the pump pulse were also investigated with the help of a rudimentary home-built time-of-flight mass spectrometer (TOF-MS) installed into the transient-absorption chamber. By placing a positive potential on the TAS target body and positioning it 1~mm above the laser interaction region, the bottom of the TAS target doubles as a repeller that accelerates the strong-field ionized molecular fragments towards an ion detector. This detector takes the form of a Photonis MegaSpiraltron channeltron placed 16~cm below the interaction region. The field-free drift region of the TOF is achieved by placing a grounded mesh just above the channeltron and using a small diameter vacuum chamber, which through the Faraday-cage-effect, causes the potential of the TAS body to decay rapidly. A reflection mesh is used to redirect the ions into the channeltron once they have passed through the grounded mesh. 

\subsection{Extended analysis of TOF-MS results}

\begin{figure}[h]
    \centering
    \includegraphics[width=0.75\textwidth]{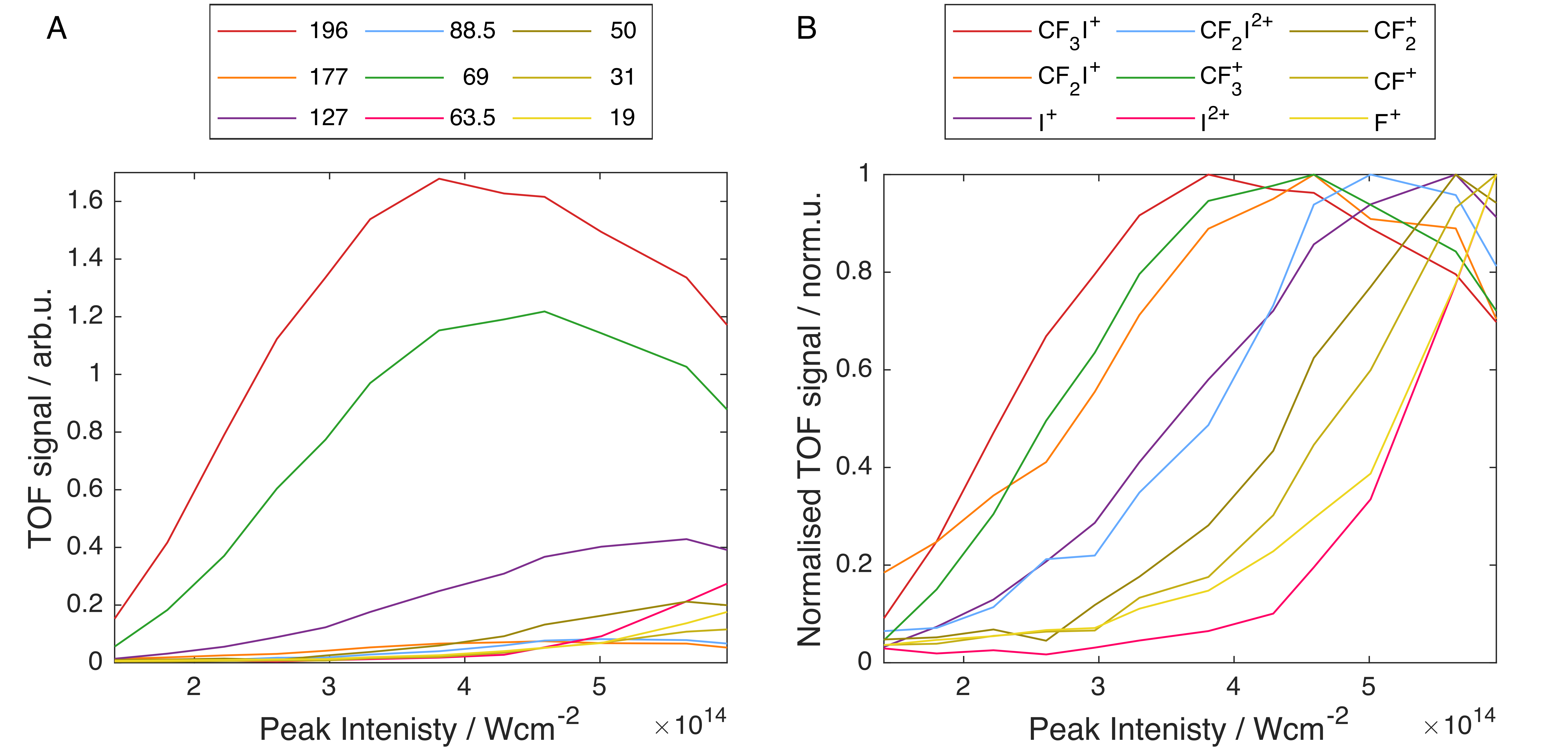}
    \caption{Intensity-resolved mass spectrometry on strong-field-ionized CF$_3$I. \textbf{A} The integrated ion yield of the different fragments as a function of ionizing-pulse intensity labeled according to their m/z ratio. \textbf{B} The integrated ion yields normalized to their maximum absolute yields, labeled with the chemical formula of the fragment.}
    \label{fig:tof}
\end{figure}

The intensity-resolved results of the TOF-MS measurements are shown in Fig.~\ref{fig:tof}. The peaks labeled in the TOF spectrum in Fig.~3D of the main text are integrated and plotted as a function of the laser peak intensity in Fig.~\ref{fig:tof}A and B, with the latter being normalized to their highest abundance. From the un-normalized results, we can see that the three most abundant fragments generated by the strong-field ionization are CF$_3$I$^+$, CF$_3^+$ and, at intensities above $3\times 10^{14}$ Wcm$^{-2}$, I$^+$. These are the expected fragmentation products of the $\tilde{\mathrm{X}}$, $\tilde{\mathrm{A}}$ and $\tilde{\mathrm{B}}$ states, respectively. Furthermore, the results show that the CF$_3$I$^+$ $\tilde{\mathrm{X}}$ state and CF$_3^+$ $\tilde{\mathrm{A}}$ state signals experience saturation effects and reach maxima at an intensity around $4\times 10^{14}$ Wcm$^{-2}$. At these intensities, the few-cycle pump pulse not only depletes the CF$_3$I ground state population, but also starts to doubly ionize the sample, reducing the yield of the energetically lower-lying cationic states. These doubly ionized molecules are unstable and fragment by distributing the charge and breaking both the C-I and C-F bonds, producing the CF$^+$, CF$_2^+$ and even the F$^+$ fragments, or by concentrating the charge and producing the I$^{2+}$ dication \cite{Crane2022}. Intensities of 2 and $3\times 10^{14}$ Wcm$^{-2}$ have therefore been used for our ATAS measurements to maximize the difference in the relative signal of the $\tilde{\mathrm{B}}$ state dynamics while minimizing the contributions of higher-energy channels.

The CF$_2$I$^{2+}$ dication shows a similar intensity dependence as the I$^+$ cation, while exhibiting a 20-fold lower absolute fragmentation yield. This correlation with I$^+$ is peculiar as in previous strong-field ionization studies \cite{Crane2022} using a 40~fs, 800~nm laser pulse, CF$_2$I$^{2+}$ was determined to only originate from doubly ionized molecules. The absolute fragmentation was, however, far lower in the case of the narrower bandwidth 40~fs pulse, compared to our octave-spanning few-cycle pulse. The higher yield when using a broader spectrum indicates that the CF$_2$I$^{2+}$ fragment may originate from an additional resonant electronic excitation of the $\tilde{\mathrm{B}}$ state into the $\tilde{\mathrm{C}}$ or $\tilde{\mathrm{D}}$ states. These states are described by the ionization of an electron from non-fully-symmetric combinations of the fluorine F-orbitals, and are therefore capable of driving dynamics that break C-F bonds.

We note that this correlation-based analysis of the ion yields is performed on results obtained from a TOF-MS that was not purposefully built for the absolute determination of ion yields, but rather, the identification of ions. Nevertheless, we have demonstrated here that the insight gained from exploring the intensity-dependence TOF-MS is very powerful for determining coupled channels in strong-field ionization, even in the absence of a precise intensity calibration.

\subsection{ATAS data acquisition and processing}

A detailed description of the data-acquisition procedure is provided in section 2.1.9 of reference \cite{matselyukh2023attosecond}. The 
ATAS results shown in Figs.~3-5 of the main text were acquired using the pulsed-nozzle-fed 3~mm tube target. Reference measurements at each delay were acquired by turning off the pulsed nozzle. A delay range of 100~fs was investigated using 500~as steps. The exposure time for each acquisition was set to 4 seconds and the measurement repeated three times (producing three `scans').

To filter out the shot-to-shot correlated fluctuations in the XUV spectrum of the probe pulse, a singular-value-decomposition- (SVD) based method was used. This methods was introduced in \cite{matselyukh22a} and formally presented in section 2.3 of \cite{matselyukh2023attosecond}. The method decomposes the transient-absorption results of the three scans into the singular vectors. The reproducibility of the delay vector is then used to assess whether the vector describes true transient signals, or correlated fluctuations in the XUV spectrum. Those vectors that are not reproducible between scans fall into the latter category and are removed from the transient absorption results by having their singular values set to zero. 

\begin{figure} [h]
    \centering
    \includegraphics[width = \textwidth]{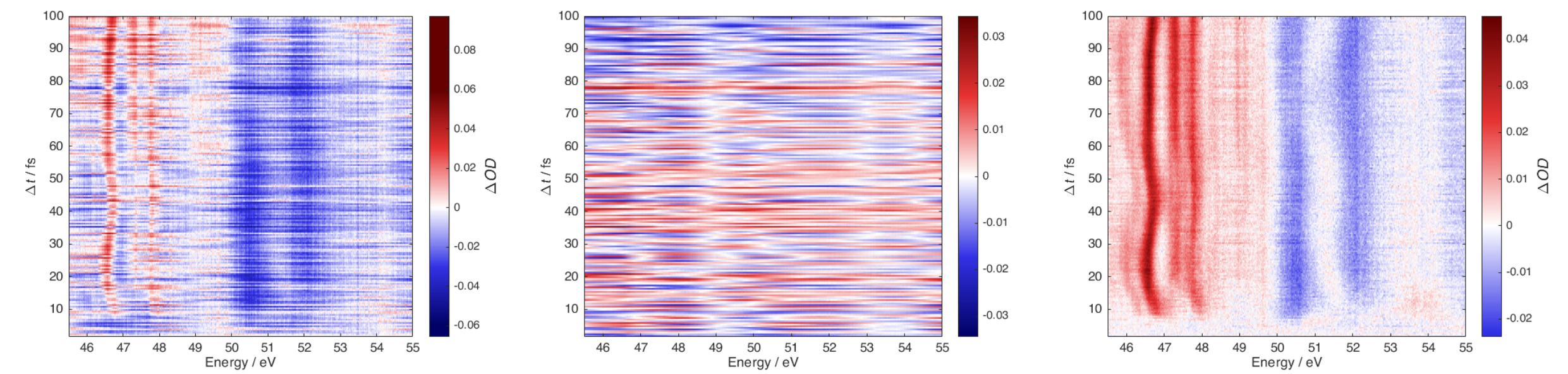}
    \caption{\label{fig:SVD}The result of the SVD-based filtration. The left panel shows the measured scan-averaged transient absorption. The middle column shows the signals originating from fluctuations in the XUV spectrum. The right-hand column shows difference between the first two panels: the SVD-filtered ATAS results.}
    \label{fig:enter-label}
\end{figure}

This filtration methodology requires the user to specify two parameters; the cut-off SVD index for filtration (beyond which the singular vectors are not filtered and not removed) and the cut-off reproducibility (the value of the reproducibility parameter below which a singular vector is removed). For the results presented in Figs.~4 and 5 of the main text, the cut-off index is set to 12 and the reproducibility cut-off to 0.1. The result of the SVD-based filtering is presented in Fig.~\ref{fig:SVD}.

\subsection{Fitting procedure for experimental results}

To maximize the robustness of the fits, we take every camera pixel which falls into the 47.25--47.36~eV and 53.50--54.30~eV photon-energy ranges to be an independent measurement of the diabatic population of the $\tilde{\mathrm{B}}$ and $\tilde{\mathrm{E}}$ states, respectively, and employ a bisquare-weighted regression. A smaller spectral range is chosen for the $\tilde{\mathrm{E}}$ state than in Fig.~4A of the main text due to the fact that the $\tilde{\mathrm{X}}_{3/2}$ state absorption extends slightly above 47.09~eV for $\Delta t <8$~fs. The data and resulting fits are shown in Fig.~5A of the main text.

\section{Computational Details}
\subsection{Geometry of the Optimized Ground State Minimum}

The minimum geometry of the ground state of \ce{CF3I} was optimized with Gaussian 16~\cite{Frisch2016-lk} using density functional theory (DFT) with the $\omega$B97X-D functional~\cite{Chai2008-eo} and the basis set 6-311G~\cite{Glukhovtsev1995-yv, Krishnan1980-fy}. The basis set was taken from the Basis Set Exchange (BSE)~\cite{Pritchard2019-hv,Schuchardt2007-wy,Feller1996-gc}. Table~\ref{tab:geom_min} shows the $xyz$ coordinates of the optimized geometry in {\AA}ngstr\"om.

 \begin{table}[h!]
    \caption{\label{tab:geom_min}Geometry of the $\omega$B97X-D/6-311G optimized ground state minimum.}
	\vspace{5mm}
	\centering
	\begin{tabular}{ccccc}
        \toprule
Number & Element & $x$ (\si{\angstrom}) & $y$ (\si{\angstrom}) & $z$ (\si{\angstrom}) \\ \midrule
1 & \ce{C} & \tablenum[table-format=-1.6]{-0.000000} & \tablenum[table-format=-1.6]{ 0.000000} & \tablenum[table-format=-1.6]{-0.004017} \\
2 & \ce{I} & \tablenum[table-format=-1.6]{-0.000000} & \tablenum[table-format=-1.6]{ 0.000000} & \tablenum[table-format=-1.6]{ 2.158913} \\
3 & \ce{F} & \tablenum[table-format=-1.6]{ 1.242440} & \tablenum[table-format=-1.6]{ 0.000000} & \tablenum[table-format=-1.6]{-0.467045} \\
4 & \ce{F} & \tablenum[table-format=-1.6]{-0.621220} & \tablenum[table-format=-1.6]{ 1.075985} & \tablenum[table-format=-1.6]{-0.467045} \\
5 & \ce{F} & \tablenum[table-format=-1.6]{-0.621220} & \tablenum[table-format=-1.6]{-1.075985} & \tablenum[table-format=-1.6]{-0.467045} \\
 	    \bottomrule
	\end{tabular}
\end{table}

\subsection{Validation of the Active Spaces for \ce{CF3I} and \ce{CF3I+}}

\begin{figure}[h!]
    \centering
    \includegraphics[width=0.5\textwidth]{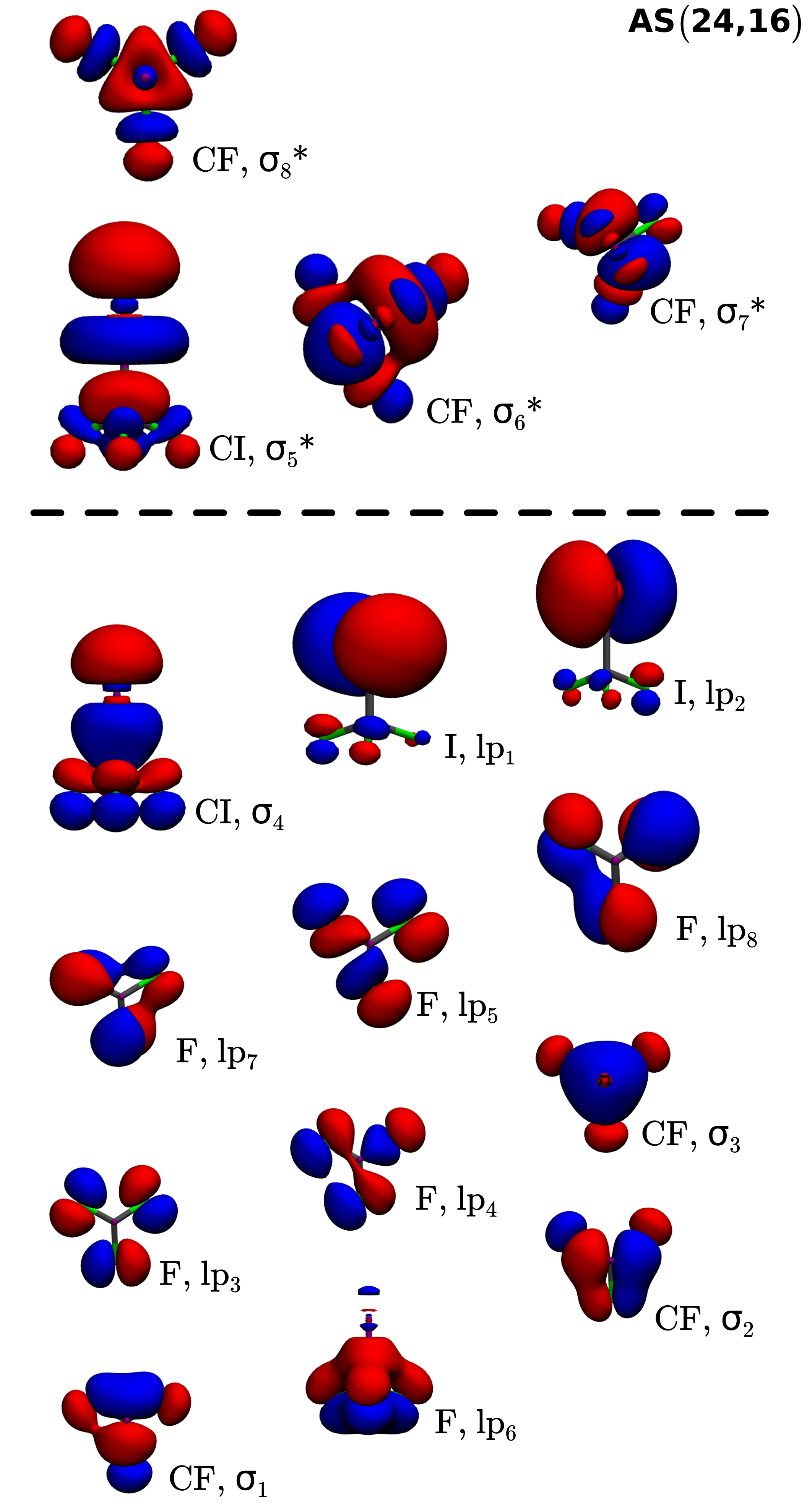}
    \caption{\label{fig:cf3i_as_big}CASSCF molecular orbitals included in the active space AS(24,16) of trifluoroiodomethane, obtained using the ANO-RCC-VDZP basis set at the $\omega$B79X-D/6-311G optimized ground state minimum geometry. Orbitals are rendered with an isovalue of $0.04$. Orbitals $\sigma_{4}$, $\sigma^{*}_{5}$, $lp_{1}$, $lp_{2}$ and $lp_{6}$ are shown in a side view, whereas the rest are shown from the top along the \ce{C\bond{1}I} bond.}
\end{figure}

In order to sufficiently describe the valence excited states of \ce{CF3I} and \ce{CF3I+}, three different active spaces (ASs) were tested. All calculations were carried out with the \textsc{OpenMolcas}~\cite{Fdez_Galvan2019-gt,Aquilante2020-hx} program package using the ANO-RCC~\cite{Roos2008-ud,Roos2005-qs,Roos2005-qj,Roos2004-ry,Roos2004-ze} basis set, contracted to VDZP quality (ANO-RCC-VDZP). The smallest AS, forming the common basis for the two larger ASs, included 12 electrons in 10 orbitals [AS(12,10)]. It consisted of the carbon-iodine bond ($\sigma_{4}$, $\sigma^{*}_{5}$), both iodine lone-pair orbitals $lp_{1}$ and $lp_{2}$ as well as the three carbon-fluorine bonds ($\sigma_{1}$, $\sigma^{*}_{6}$, $\sigma_{2}$, $\sigma^{*}_{7}$ and $\sigma_{3}$, $\sigma^{*}_{8}$). Subsequently, the medium AS was extended by three fluorine lone-pair orbitals $lp_{3}$, $lp_{4}$ and $lp_{5}$ resulting in 18 electrons in 13 orbitals [AS(18,13)]. The large AS was extended by an additional three orbitals ($lp_{6}$, $lp_{7}$ and $lp_{8}$), including all six fluorine lone-pair orbitals [AS(24,16)]. The orbitals included in the ASs are shown in Fig.~\ref{fig:cf3i_as_big}. For the calculation of the cation, the same orbitals are included but with one electron removed, resulting in the set AS(11,10), AS(17,13) and AS(23,16).\\

As a first validation step for the three ASs, the ionization energies were compared to the experimental values taken from the study of Yates and coworkers~\cite{Yates1986-sw}. The ionization energies listed in Table~\ref{tab:ionization_energies} were calculated at the MS-CASPT2 level of theory, including the effects of spin-orbit coupling (SOC), as the difference between the GS of the neutral \ce{CF3I} and the excited states of the cation. Further, we compared the electronic character of the cationic states to calculated charge distributions taken from the same publication as listed in Table~\ref{tab:ionization_chargedistribution}. The first two states, $\tilde{X}\ (^{2}E_{\sfrac{3}{2}})$ and $\tilde{X}\ (^{2}E_{\sfrac{1}{2}})$, describe an ionization from either of the \ce{I} lone pair orbitals $lp_{1}$ and $lp_{2}$. Their ionization energies are described quite well by all three ASs. The third state, $\tilde{B}\ (^{2}A_{1})$, an ionization from the \ce{C\bond{1}I} bonding $\sigma_{4}$ orbital, is correctly described by the AS(23,16), with the other two overestimating the ionization energy. For the next three states, the electron hole is generated in the three \ce{F} lone-pair orbitals $lp_{3}$, $lp_{4}$ and $lp_{5}$. These states cannot be described by the AS(11,10), as the necessary orbitals are not included. The other two ASs correctly describe their electronic character and more or less capture the range of the ionization energy. However, the AS(17,13) does not describe the correct order of the states as, energetically, the last state $\tilde{E}$ appears between the states $\tilde{B}$ and $\tilde{C}$. For the $\tilde{E}$ state, the hole is again located mainly on the \ce{I} with the $lp_{1}$, $lp_{2}$ and $\sigma^{*}_{5}$ orbitals partially occupied. Here, only the AS(23,16) is able to correctly describe the order of the electronic states. It does fall short in matching the ionization energies of the highest states, but as all three ASs struggle in this regard, we suspect the limiting factor to be the moderate size of the basis set ANO-RCC-VDZP, as it is, for example, unable to sufficiently describe possible Rydberg-type contributions to the orbitals.

\begin{table}[h!]
	\caption{\label{tab:ionization_energies}Ionization energies $\Delta E$ at the FC point for the first seven states of the \ce{CF3I+} cation in \unit{\electronvolt} at the MS-CASPT2 level of theory, including spin-orbit couplings. The energies were calculated as the difference of the GS energy of neutral species and the energies of the cationic states $\tilde{X}$-$\tilde{E}$. For the electronic character of the states, the partially occupied orbitals of the dominant configuration-interaction vector based on the calculation with the AS(23,16) are listed. The experimental ionization energies are taken from the work of Yates and coworkers~\cite{Yates1986-sw}}
	\centering
	\vspace{5mm}
	\begin{tabular}{cccccc}
	\toprule
 & &\multicolumn{4}{c}{Ionization energies $\Delta E$ (\unit{\electronvolt})} \\ \cmidrule(lr){3-6}
State & Character & AS(11,10) & AS(17,13) & AS(23,16) & Exp.~\cite{Yates1986-sw} \\ \midrule
$\tilde{X}\ (^{2}E_{\sfrac{3}{2}})$ & $lp_{1}$ & 10.74 & 10.49 & 10.54 & 10.45 \\
$\tilde{X}\ (^{2}E_{\sfrac{1}{2}})$ & $lp_{2}$ & 11.36 & 11.11 & 11.12 & 11.18 \\
$\tilde{A}\ (^{2}A_{1})$ & $\sigma_{4}$ & 14.07 & 13.74 & 13.30 & 13.25 \\
$\tilde{B}\ (^{2}A_{2})$ & $lp_{3}$ & - & 15.15 & 15.22 & 15.56 \\
$\tilde{C}\ (^{2}E)$ & $lp_{4}$ & - & 17.01 & 15.92 & 16.32 \\
$\tilde{D}\ (^{2}E)$ & $lp_{5}$ & - & 17.02 & 16.27 & 16.32 \\
$\tilde{E}\ (^{2}A_{1})$ & $lp_{1,2}$, $\sigma^{*}_{5}$ & 16.33 & 16.18 & 16.80 & 17.28 \\
	\bottomrule
	\end{tabular}
\end{table}

\begin{table}[h!]
	\caption{\label{tab:ionization_chargedistribution}Calculated charge distribution of the first seven cationic states of \ce{CF3I+} taken from the work of Yates and coworkers~\cite{Yates1986-sw}. ‘Out’ and ‘In’ are the percentage charge of the outersphere and intersphere regions, respectively. The column ‘Sum’ is not taken from the reference, but is an interpretation of the data. For our calculations the partially occupied orbitals of the transition with the largest configuration-interaction weight are shown as an approximation for the charge distribution.}
	\centering
	\vspace{5mm}
	\begin{tabular}{cccccccccc}
	\toprule
 & \multicolumn{6}{c}{Charge distribution [\unit{\percent}]~\cite{Yates1986-sw}} & \multicolumn{3}{c}{Partially occupied orbitals} \\ \cmidrule(lr){2-7} \cmidrule(lr){8-10}
State & Out & C & F & I & In & Sum & AS(11,10) & AS(17,13) & AS(23,16) \\ \midrule
$\tilde{X}\ (^{2}E_{\sfrac{3}{2}})$ & 4.2 & 0.1 & 1.1 & 77.6 & 17.0 & \ce{I+} & $lp_{1}$ & $lp_{1}$ & $lp_{1}$ \\
$\tilde{X}\ (^{2}E_{\sfrac{1}{2}})$ & 4.2 & 0.1 & 1.1 & 77.6 & 17.0 & \ce{I+} & $lp_{2}$ & $lp_{2}$ & $lp_{2}$ \\
$\tilde{A}\ (^{2}A_{1})$ & 1.9 & 27.9 & 18.0 & 44.2 & 8.0 & Non-local & $\sigma_{4}$ & $\sigma_{4}$ & $\sigma_{4}$ \\
$\tilde{B}\ (^{2}A_{2})$ & 0.6 & 0.0 & 80.5 & 0.0 & 18.9 & \ce{F+} & $lp_{1,2}$, $\sigma^{*}_{5}$ & $lp_{3}$ & $lp_{3}$ \\
$\tilde{C}\ (^{2}E)$ & 0.7 & 0.3 & 78.4 & 0.3 & 20.3 & \ce{F+} & $lp_{1,2}$, $\sigma^{*}_{5}$ & $lp_{1,2}$, $\sigma^{*}_{5}$ & $lp_{4}$ \\
$\tilde{D}\ (^{2}E)$ & 2.2 & 0.7 & 76.9 & 0.1 & 20.2 & \ce{F+} & $lp_{1,2}$, $\sigma^{*}_{5}$ & $lp_{4}$ & $lp_{5}$ \\
$\tilde{E}\ (^{2}A_{1})$ & 1.9 & 1.0 & 36.2 & 49.7 & 11.2 & Non-local & - & $lp_{5}$ & $lp_{1,2}$, $\sigma^{*}_{5}$ \\
	\bottomrule
	\end{tabular}
\end{table}

\begin{figure}[htbp]
    \centering
    \includegraphics[width=\textwidth]{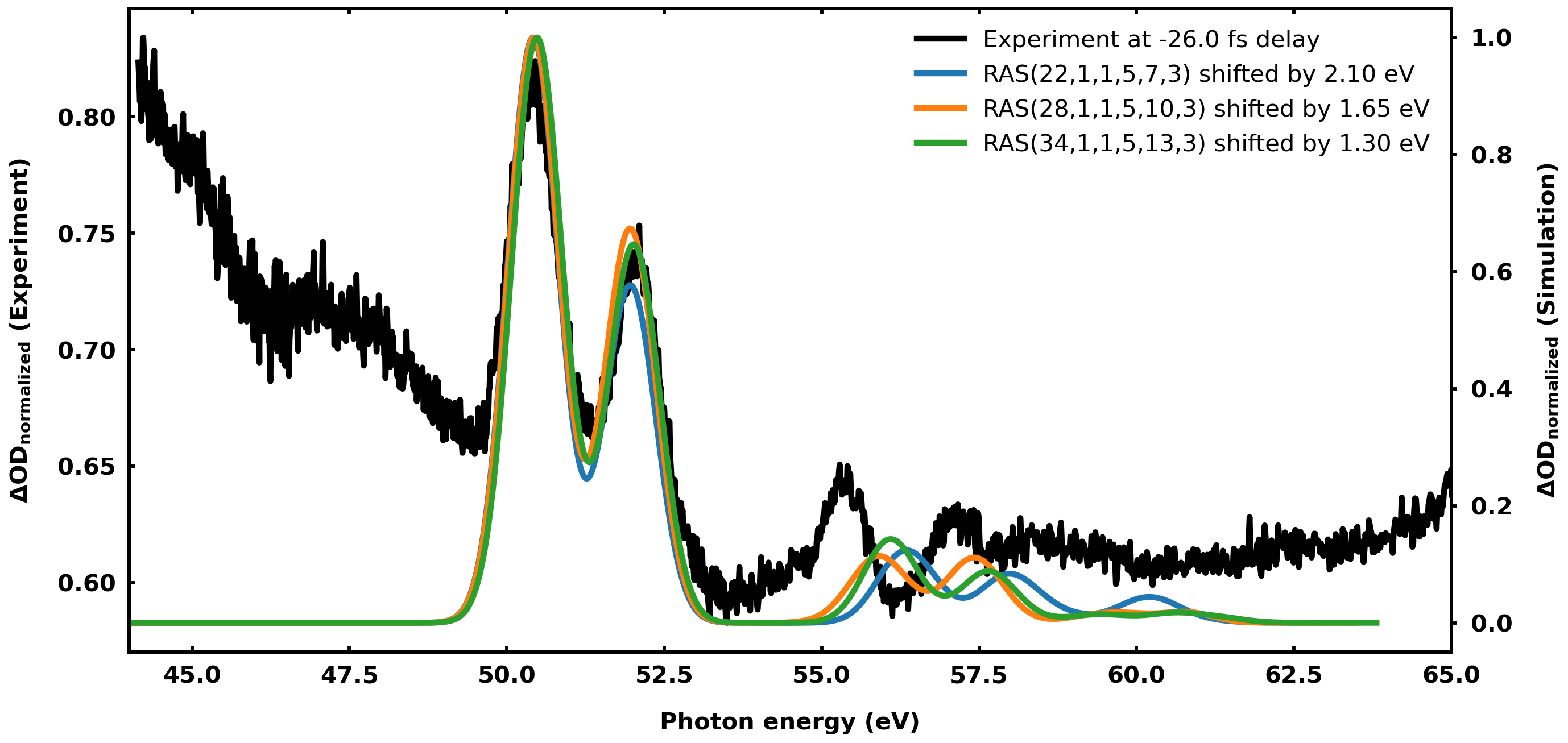}
    \caption{\label{fig:as_benchmark_gs_spectrum}GS spectra of neutral \ce{CF3I}, simulated for the three different ASs. The calculated peaks are broadened by applying a Gaussian with $\sigma=\qty{0.4}{\electronvolt}$. In order to match the experimental signal shown in black, the excitation energies needed to be shifted by different amounts.}
\end{figure}

Finally, the GS spectrum of neutral \ce{CF3I} was simulated for all three ASs following the procedure discussed in the manuscript and section~\ref{sec:xas_setup} and compared with the experimental signal, as shown in Fig.~\ref{fig:as_benchmark_gs_spectrum}. Here, all three spectra show the pronounced double-peak structures of the experiment at \qtylist{50.4;52.1}{\electronvolt} and at \qtylist{55.4;57.1}{\electronvolt}. But depending on the size of the AS, a different shift of the excitation energies needed to be applied. Shifting to the highest peak in the experiment at \qty{50.4}{\electronvolt}, resulted in values of \qtylist{2.00;1.60;1.30}{\electronvolt} for the spectra of RAS($22,1,1;5,7,3$), RAS($28,1,1;5,10,3$) and RAS($34,1,1;5,13,3$), receptively. In general, the prominent doublet at \qtylist{50.4;52.1}{\electronvolt} is described quite well by all three ASs. With the RAS($34,1,1;5,13,3$) it was even possible to reproduce the correct intensity distribution of the said doublet. For the second doublet at \qtylist{55.4;57.1}{\electronvolt}, all three spectra are energetically off by about \qtyrange{0.7}{0.8}{\electronvolt}. Here, it is again possible that the size of the basis set ANO-RCC-VDZP prohibits a perfect match of this feature. However, considering the fact that calculations with a larger basis set were computationally infeasible and the otherwise excellent agreement with the experimental spectrum, we are convinced that the ASs (24,16) and (23,16) are adequate to describe the valence space and core-excited states of neutral and ionic \ce{CF3I}.

\begin{figure}[htbp]
    \centering
    \includegraphics[width=\textwidth]{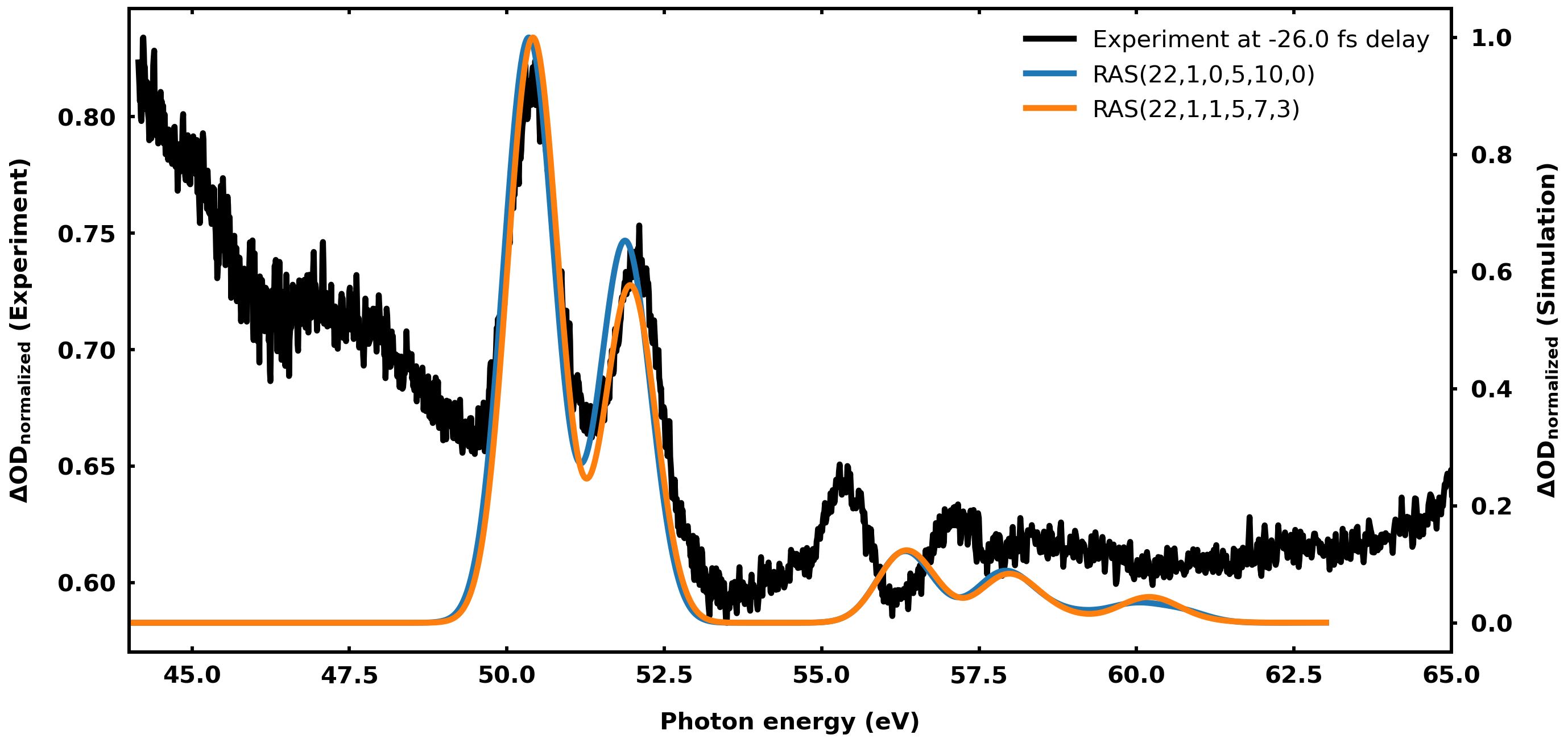}
    \caption{\label{fig:as_ras3_trick}GS spectra of neutral \ce{CF3I}, simulated for the AS(12,10). In both cases, the excitation energies were shifted by \qty{2.00}{\electronvolt} to match the experimental signal shown in black and broadened by applying a Gaussian with $\sigma=\qty{0.4}{\electronvolt}$. The spectrum where all orbitals are included in the RAS2, is shown in blue. And the one where the AS is split up between the RAS3 and RAS2 sub spaces is shown in orange.}
\end{figure}

\newpage
\subsection{Methodology for obtaining the XAS}
\label{sec:xas_setup}
The general procedure, on how to calculate an XUV absorption spectrum (further abbreviated as only XAS) based on the RASSCF/RASPT2 ansatz was introduced in detail in our previous study on the ultrafast strong-field dissociation of vinyl bromide~\cite{Rott2021-vb}. So, here we only mention the aspects of the calculations that differ from the introduced procedure.\\

\begin{figure}[h!]
    \centering
    \includegraphics[width=0.6\textwidth]{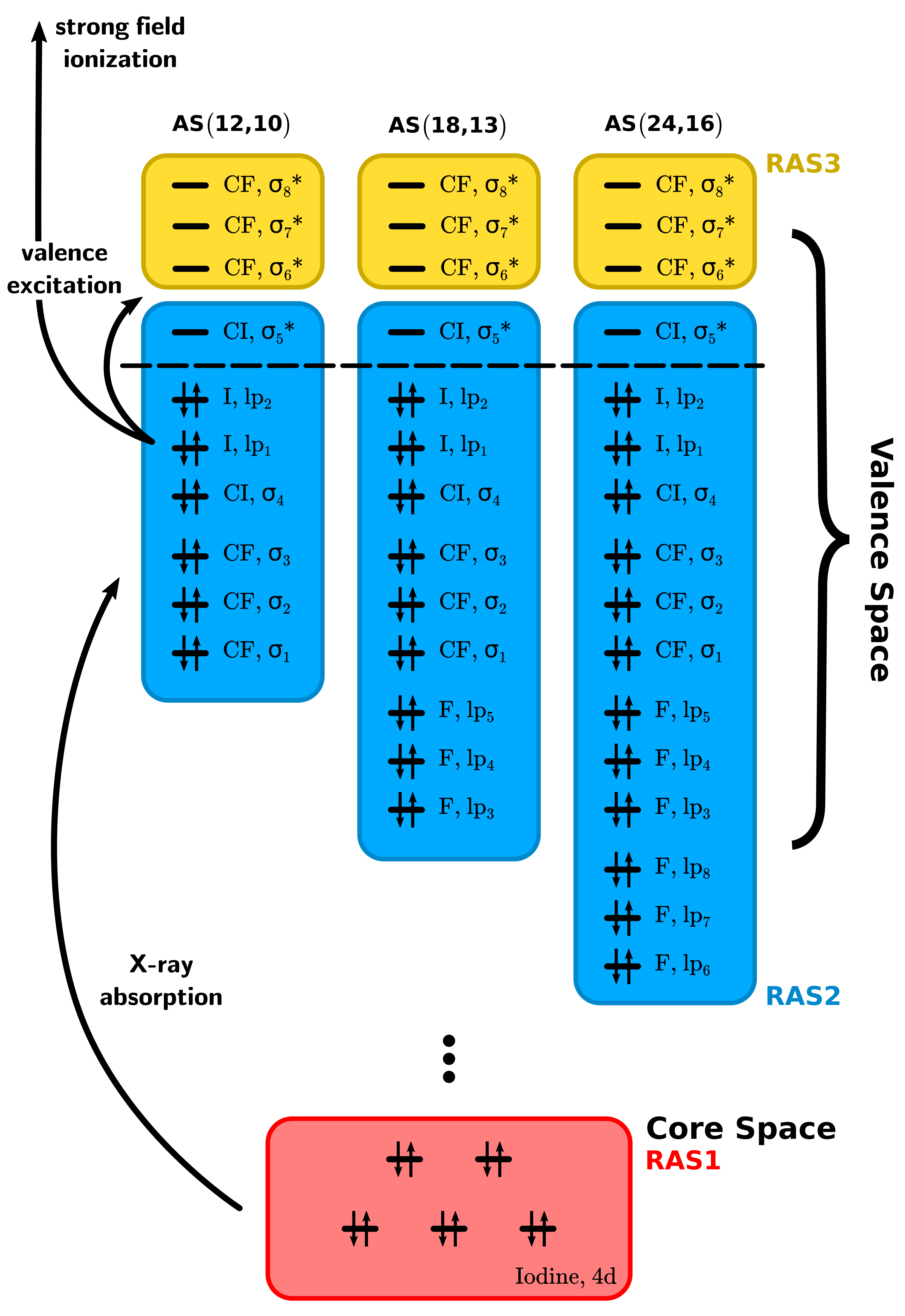}
    \caption{\label{fig:xas_diagram}Diagram of the three active spaces RAS($22,1,1;5,7,3$), RAS($28,1,1;5,10,3$) and RAS($34,1,1;5,13,3$) tested for the simulation of the XAS of \ce{CF3I}. They are based on the previously introduced active space AS(12,10), AS(18,13) and AS(24,26).}
\end{figure}

Based on the AS(24,16) shown in Fig.~\ref{fig:cf3i_as_big}, the necessary RAS sub-spaces were set up for the calculation of the core-excited states. In general the sub-spaces can be systematically labeled RAS($n,l,m;i,j,k$), where, $i$, $j$ and $k$ are the number of orbitals in the RAS1, RAS2, and RAS3 sub-spaces, respectively, $n$ is the total number of electrons in the AS, $l$ the maximum number of holes allowed in the RAS1, and $m$ the maximum number of electrons allowed in RAS3. As the experiment probed at the $N_{4,5}$ edge of iodine, its five $4d$ orbitals were included in the RAS1. To reduce computational costs, the original AS(24,16) was split between the RAS2 and RAS3 sub-spaces. The three virtual orbitals of the carbon-fluorine bonds ($\sigma^{*}_{6}$, $\sigma^{*}_{7}$ and $\sigma^{*}_{8}$) made up the RAS3 and a single excitation into them was allowed. The remaining 13 orbitals were included in the RAS2 sub-space, resulting in the RAS($34,1,1;5,13,3$) illustrated in Fig.~\ref{fig:xas_diagram}. To judge whether this influenced the simulated spectrum, we compared the GS spectrum of the 'full' AS with the approximation of splitting the AS. As this benchmark was performed for the AS(12,10), the resulting RAS are RAS($22,1,0;5,10,0$) and RAS($22,1,1;5,7,3$). Both calculated spectra, shifted by \qty{2.00}{\electronvolt} to match the experimental signal, are shown in Fig.~\ref{fig:as_ras3_trick}. The spectra are nearly identical, with the only difference in the intensity of the second peak of the doublet at \qtylist{50.4;52.1}{\electronvolt}, which is slightly lower, when the AS is split between the two RAS sub-spaces. In summary, since the splitting of the AS does not significantly change the simulated spectrum, and one can safely use this approximation, at least in the case of trifluoroiodomethane. But the total calculation time, necessary to arrive at the final spectrum, could be cut down drastically. The initial calculation time of over eight days for RAS($22,1,0;5,10,0$) could be reduced to about four and a half hours by utilizing the RAS3 sub-space. This speed-up made simulations with the AS(24,16) possible, as they could be completed in about three days. For the original setup of just using the RAS2 sub spaces, these calculations were not feasible as a total calculation time of over three months was estimated.

\subsection{PECs and XAS of cationic Trifluoroiodomethane}

\begin{figure}[h!]
    \centering
    \includegraphics[width=\textwidth]{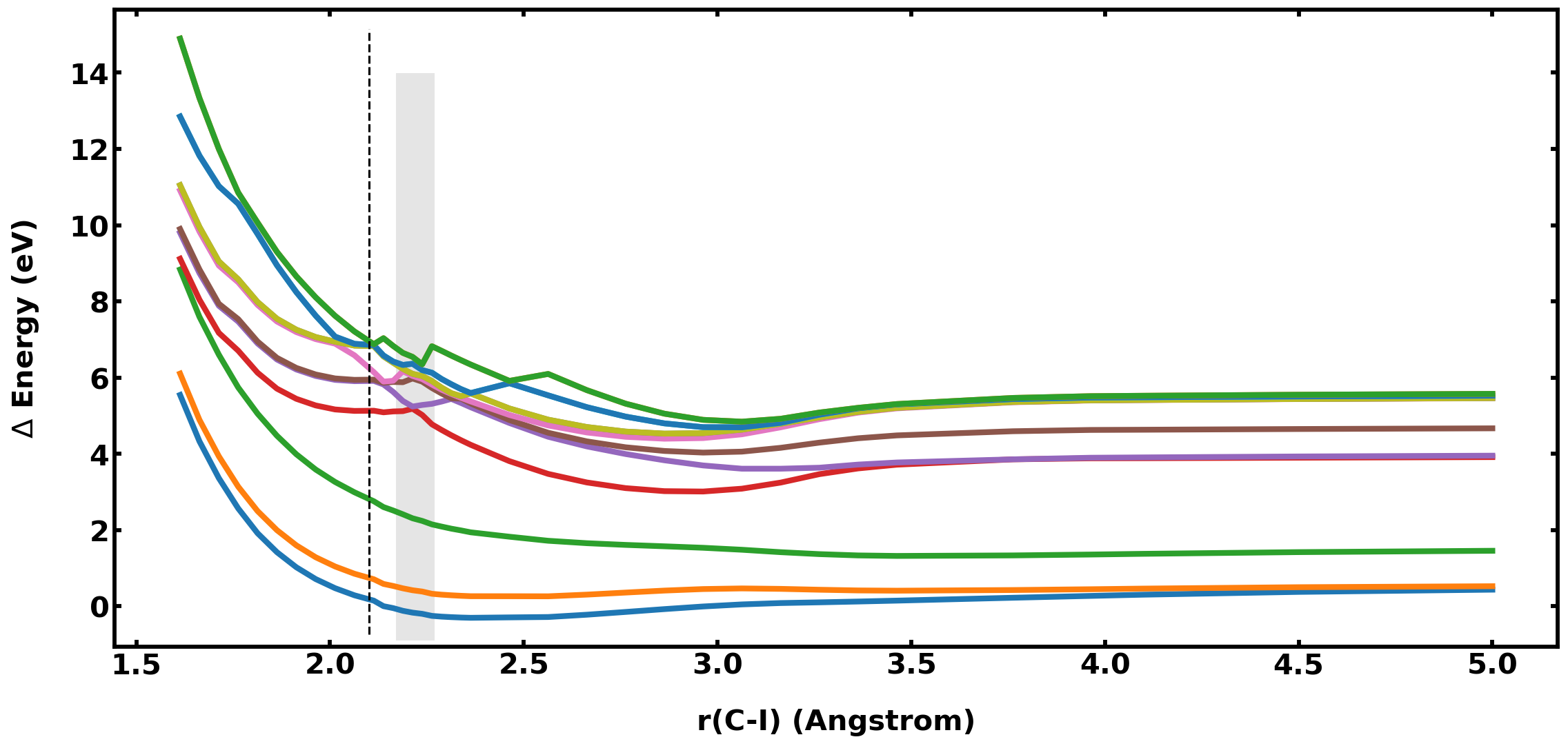}
    \caption{\label{fig:cf3i_relaxed_scan}Relaxed scan of the \ce{C\bond{1}I} bond. The states are color-coded corresponding to their adiabatic state order. The FC region is indicated by the black dotted line and the gray box highlights the region of strong coupling.}
\end{figure}

To further analyze the dissociation dynamics of the cationic trifluoroiodomethane after excitation, we performed a relaxed scan of the \ce{C\bond{1}I} bond. The relaxed scan was performed with OpenMolcas utilizing the state tracking feature (keyword \textit{TRACk}) of the \textit{SLAPAF} program, where the character of a specific state can be followed throughout a geometry optimization. In our case, we chose to follow the character of the $\tilde{X}$ state at the Frank-Condon region. At each step in the scan, the geometry was optimized at the SA11-CASSCF(17,13)/ANO-RCC-VDZP level of theory. The \ce{C\bond{1}I} bond length was scanned from \SI{1.6}{\angstrom} to \SI{5.0}{\angstrom} with an initial step size of \SI{0.05}{\angstrom} but a bond length of \SI{2.3}{\angstrom} it was increased to \SI{0.1}{\angstrom}. In the Frank-Condon (FC) region between \SIrange{2.0}{2.3}{\angstrom} the steps size was reduced to \SI{0.025}{\angstrom}. The resulting PECs are plotted in Fig.~\ref{fig:cf3i_relaxed_scan}.

The electronic character of all relevant states, the partially occupied orbitals of the dominant configuration-interaction vector and its weight based on the calculation with the AS(23,16) along the relaxed scan are listed in Table~\ref{tab:electronic_character}.

\begin{table}[h!]
\caption{\label{tab:electronic_character}The electronic character of the seven relevant states at the optimized geometries of the relaxed scan of the \ce{C\bond{1}I} bond between \SI{1.962}{\angstrom} and \SI{2.462}{\angstrom}. For the electronic character of the states, the partially occupied orbitals of the dominant configuration-interaction vector and its weight in \unit{\percent} based on the calculation with the AS(23,16) are listed.}
	\centering
	\vspace{5mm}
	\begin{tabular}{ccccccccccc}
 	\toprule
  & \multicolumn{10}{c}{electronic character at $R_{CI}$ (\unit{\angstrom})} \\ \cmidrule(lr){2-11}
\# & \multicolumn{2}{c}{1.962} & \multicolumn{2}{c}{2.012} & \multicolumn{2}{c}{2.062} & \multicolumn{2}{c}{2.112} & \multicolumn{2}{c}{2.137} \\ \midrule
1 & $lp_{1}$     & 52.8 & $lp_{1}$     & 52.0 & $lp_{1}$     & 92.0 & $lp_{1}$     & 91.2 & $lp_{1}$                     & 94.5 \\
2 & $lp_{2}$     & 53.0 & $lp_{2}$     & 52.3 & $lp_{2}$     & 92.6 & $lp_{2}$     & 92.2 & $lp_{2}$                     & 94.6 \\
3 & $\sigma_{4}$ & 92.3 & $\sigma_{4}$ & 92.1 & $\sigma_{4}$ & 92.0 & $\sigma_{4}$ & 91.6 & $\sigma_{4}$                 & 92.3 \\
4 & $lp_{3}$     & 89.9 & $lp_{3}$     & 89.3 & $lp_{3}$     & 89.0 & $lp_{3}$     & 88.4 & $lp_{3}$                     & 84.8 \\
5 & $lp_{4}$     & 65.3 & $lp_{4}$     & 62.3 & $lp_{4}$     & 60.0 & $lp_{4}$     & 58.8 & $lp_{4}$                     & 83.5 \\
6 & $lp_{5}$     & 67.3 & $lp_{5}$     & 64.9 & $lp_{5}$     & 62.2 & $lp_{5}$     & 60.0 & $lp_{5}$                     & 83.2 \\
7 & $lp_{7}$     & 67.2 & $lp_{7}$     & 64.8 & $lp_{7}$     & 61.9 & $lp_{7}$     & 59.4 & $lp_{1,2}$, $\sigma^{*}_{5}$ & 87.0 \\
	\bottomrule
 \\ 
 \toprule
  & \multicolumn{10}{c}{electronic character at $R_{CI}$ (\unit{\angstrom})} \\ \cmidrule(lr){2-11}
\# & \multicolumn{2}{c}{2.162} & \multicolumn{2}{c}{2.187} & \multicolumn{2}{c}{2.212} & \multicolumn{2}{c}{2.237} & \multicolumn{2}{c}{2.262} \\ \midrule
1 & $lp_{1}$                     & 94.6 & $lp_{1}$                     & 95.0 & $lp_{1}$                     & 95.0 & $lp_{1}$                                 & 94.2 & $lp_{1}$                                 & 94.5 \\
2 & $lp_{2}$                     & 94.1 & $lp_{2}$                     & 95.1 & $lp_{2}$                     & 95.0 & $lp_{2}$                                 & 94.3 & $lp_{2}$                                 & 94.5 \\
3 & $\sigma_{4}$                 & 92.1 & $\sigma_{4}$                 & 91.8 & $\sigma_{4}$                 & 91.9 & $\sigma_{4}$                             & 91.8 & $\sigma_{4}$                             & 89.1 \\
4 & $lp_{3}$                     & 84.4 & $lp_{3}$                     & 80.4 & $lp_{1,2}$, $\sigma^{*}_{5}$ & 87.6 & $lp_{1,2}$, $\sigma^{*}_{5}$             & 87.2 & $lp_{1,2}$, $\sigma^{*}_{5}$             & 91.1 \\
5 & $lp_{1,2}$, $\sigma^{*}_{5}$ & 86.4 & $lp_{1,2}$, $\sigma^{*}_{5}$ & 86.7 & $lp_{3}$                     & 80.8 & $lp_{3}$                                 & 76.5 & $lp_{1}$, $\sigma_{4}$, $\sigma^{*}_{5}$ & 57.6 \\
6 & $lp_{4}$                     & 82.6 & $lp_{4}$                     & 74.4 & $\sigma^{*}_{5}$             & 29.5 & $lp_{1}$, $\sigma_{4}$, $\sigma^{*}_{5}$ & 51.0 & $lp_{2}$, $\sigma_{4}$, $\sigma^{*}_{5}$ & 53.0 \\
7 & $lp_{5}$                     & 82.6 & $lp_{1,2}$, $\sigma^{*}_{5}$ & 81.4 & $lp_{1,2}$, $\sigma^{*}_{5}$ & 67.8 & $lp_{2}$, $\sigma_{4}$, $\sigma^{*}_{5}$ & 44.2 & $\sigma^{*}_{5}$                         & 38.0 \\
	\bottomrule
\\
 \toprule
  & \multicolumn{10}{c}{electronic character at $R_{CI}$ (\unit{\angstrom})} \\ \cmidrule(lr){2-11}
\# & \multicolumn{2}{c}{2.287} & \multicolumn{2}{c}{2.312} & \multicolumn{2}{c}{2.337} & \multicolumn{2}{c}{2.362} & \multicolumn{2}{c}{2.462} \\ \midrule
1 & $lp_{1}$                                 & 94.2 & $lp_{1}$                                 & 94.0 & $lp_{1}$                                 & 93.8 & $lp_{1}$                                 & 93.6 & $lp_{1}$                                 & 92.9 \\
2 & $lp_{2}$                                 & 94.2 & $lp_{2}$                                 & 94.0 & $lp_{2}$                                 & 93.8 & $lp_{2}$                                 & 93.7 & $lp_{2}$                                 & 92.9 \\
3 & $\sigma_{4}$                             & 89.0 & $\sigma_{4}$                             & 88.9 & $\sigma_{4}$                             & 88.8 & $\sigma_{4}$                             & 88.8 & $\sigma_{4}$                             & 88.6 \\
4 & $lp_{1,2}$, $\sigma^{*}_{5}$             & 91.0 & $lp_{1,2}$, $\sigma^{*}_{5}$             & 90.7 & $lp_{1,2}$, $\sigma^{*}_{5}$             & 90.6 & $lp_{1,2}$, $\sigma^{*}_{5}$             & 90.4 & $lp_{1,2}$, $\sigma^{*}_{5}$             & 90.1 \\
5 & $lp_{1}$, $\sigma_{4}$, $\sigma^{*}_{5}$ & 67.0 & $lp_{1}$, $\sigma_{4}$, $\sigma^{*}_{5}$ & 67.7 & $lp_{1}$, $\sigma_{4}$, $\sigma^{*}_{5}$ & 68.1 & $lp_{1}$, $\sigma_{4}$, $\sigma^{*}_{5}$ & 68.0 & $lp_{1}$, $\sigma_{4}$, $\sigma^{*}_{5}$ & 66.5 \\
6 & $lp_{2}$, $\sigma_{4}$, $\sigma^{*}_{5}$ & 67.5 & $lp_{2}$, $\sigma_{4}$, $\sigma^{*}_{5}$ & 68.0 & $lp_{2}$, $\sigma_{4}$, $\sigma^{*}_{5}$ & 68.4 & $lp_{2}$, $\sigma_{4}$, $\sigma^{*}_{5}$ & 68.2 & $lp_{2}$, $\sigma_{4}$, $\sigma^{*}_{5}$ & 66.5 \\
7 & $\sigma^{*}_{5}$                         & 43.3 & $lp_{1,2}$, $\sigma^{*}_{5}$             & 86.6 & $lp_{1,2}$, $\sigma^{*}_{5}$             & 90.6 & $lp_{1,2}$, $\sigma^{*}_{5}$             & 91.3 &  $lp_{1,2}$, $\sigma^{*}_{5}$            & 73.2 \\
	\bottomrule
	\end{tabular}
\end{table}

Building on top of the optimized geometries we further performed single-point MS-RASPT2 calculations including spin-orbit effects utilizing the big AS(23,16) to obtain the XAS.
To reduce the computational cost for these calculations, we put the three highest virtual orbitals ($\sigma^{*}_{6}$, $\sigma^{*}_{7}$ and $\sigma^{*}_{8}$) into the RAS3 space and allowed for a single excitation [RAS(23,0,1,0,13,3)]. The results are shown in Fig.~\ref{fig:XAS_all} as well as Fig.~3F and 4B of the main text. In these figures, an energy shift of 1.36, 1.36, 1.22, 0.98, and 0.98~eV have been applied to the $\tilde{\mathrm{X}}_{3/2}$-,  $\tilde{\mathrm{X}}_{1/2}$-,  $\tilde{\mathrm{A}}$-, $\tilde{\mathrm{B}}$- and $\tilde{\mathrm{E}}$-states, respectively. The shift is determined by adjusting the position of the calculated XAS at the longest C-I bond length to best match the measured spectra of atomic iodine radicals and cations~\cite{OSullivan1996}. As the dissociation limit of the $\tilde{\mathrm{B}}$-state is unknown, its shift is set to be the same as that of the $\tilde{\mathrm{E}}$-state.

\subsection{Spectroscopic Assignment of the ground state XAS}

In the following, the features of the static GS spectrum of neutral \ce{CF3I} are assigned at the FC geometry. In the experimental signal shown in Fig.~\ref{fig:as_benchmark_gs_spectrum}, two distinct doublets at \qtylist{50.4;52.1}{\electronvolt} and \qtylist{55.4;57.1}{\electronvolt}, in both cases with a splitting of \qty{1.70}{\electronvolt}, are visible. When only a small shift of \qty{1.30}{\electronvolt} to the excitation energies, the simulated spectrum matches the experimental one very well. The first doublet at \qtylist{50.4;52.1}{\electronvolt} can be attributed to an excitation from the \ce{I} $4d$ orbitals into the $\sigma^{*}_{5}$ anti-bonding orbital of the \ce{C\bond{1}I} bond. The splitting of \qty{1.70}{\electronvolt} arises from the SO splitting of the $4d_{\sfrac{3}{2}}$ and $4d_{\sfrac{5}{2}}$ orbitals of \ce{I}. Similarly, the second doublet at \qtylist{56.1;57.7}{\electronvolt} can be attributed to an excitation into the anti-bonding orbitals $\sigma^{*}_{6}$, $\sigma^{*}_{7}$ and $\sigma^{*}_{8}$ of the \ce{C\bond{1}F} bond. However, compared to the experiment, this doublet is shifted by about \qty{0.7}{\electronvolt} to higher energies. This can be explained by the moderate size of the basis set ANO-RCC-VDZP, which may not be able to fully describe the $\sigma^{*}_{6}$, $\sigma^{*}_{7}$ and $\sigma^{*}_{8}$ orbitals, especially if these orbitals are subject to Rydberg-type contributions. Overall, the simulated steady-state spectrum from the GS matches the experimental one very well, as the intensity distribution of the two doublets is spot-on and only a moderate shift of the excitation energies needed to be applied.

\subsection{C-I bond length dependence of XAS cross-section}

To be able to make the direct connection between experimentally measured absorptions and diabatic populations, we must establish the invariance of the XAS cross section with respect to nuclear coordinates in the vicinity of the CoIn. We do this by investigating the calculated XAS in the region of the CoIn located at $R_\mathrm{CI} = 2.21$~\AA. 

\begin{figure}[h]
	\centering
	\includegraphics[width=\textwidth]{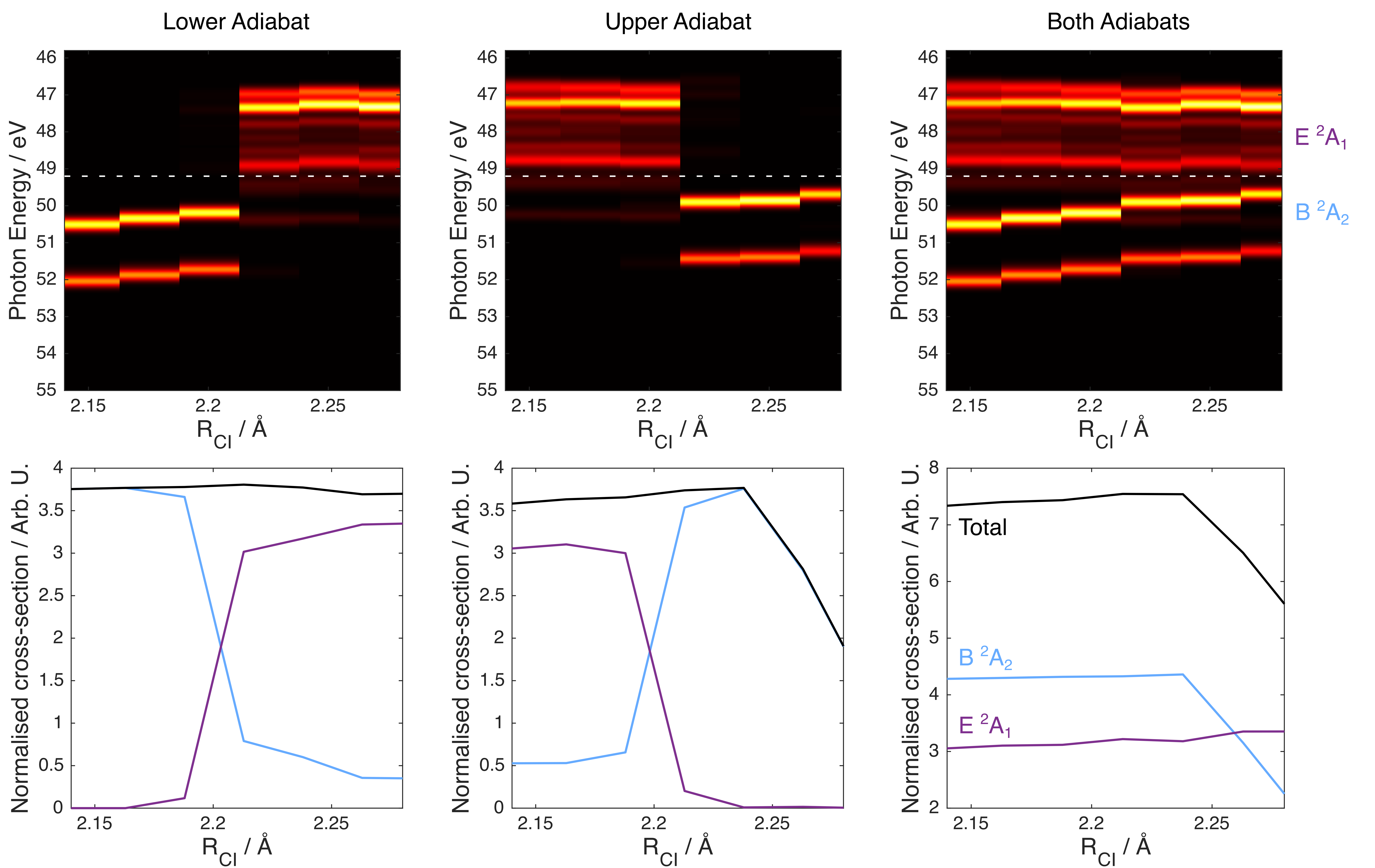}
	\caption{\label{fig:xas_crosssection}XAS as a function of C-I bond length in the region of the CoIn. The upper panels show the calculated cross sections as a function of the C-I bond length of the adiabatic states as well as the sum of the two. A dotted horizontal line indicates the energy that separates the absorption features of the $\tilde{B}$ and $\tilde{E}$ states. The lower panels show the result of integrating the upper and lower halves of the spectra shown in the upper panels. The higher energy half is multiplied by 1.85 to correct for the different integrated cross-sections of the $\tilde{B}$ and $\tilde{E}$ states.}
\end{figure}

We distinguish between the absorption of the two diabatic states by assuming that absorption at photon energies below 49.2~eV corresponds to the $\tilde{E}$ diabatic state and absorption above 49.2~eV corresponds to the $\tilde{B}$ diabatic state. This allows us to integrate over the respective photon energies and investigate the integrated cross section as a function of the C-I bond length along the two adiabats that participate in the CoIn. These are shown in Figure~\ref{fig:xas_crosssection}.

From the first two columns, we see that the absorption along the adiabatic PECs switches from one electronic character to another within one geometric step (0.3~pm) of the RASPT2 calculation as we cross the CoIn. Such a sharp switch is necessitated by the fact that coupling between the $\tilde{B}$ ($^2$A$_2$) and $\tilde{E}$ ($^2$A$_1$) states is entirely symmetry-forbidden in the C$_{\rm 3v}$ point group (see Section~S5).

The sum of the re-weighted cross-section (accounting for the different cross-section of the two diabatic states) is, however, left unchanged. The final column demonstrates the insensitivity of the absorption to the C-I bond length. The total absorption of the $\tilde{B}$ and $\tilde{E}$ states is almost constant from the FC point ($R_\mathrm{CI} = 2.14$~\AA) all the way up to 2.24~\AA\ at which point the absorption strength of the $\tilde{B}$ state starts to fall as the molecule starts to dissociate. These results therefore show that the transient absorption of the spectral regions attributed to the $\tilde{B}$ and $\tilde{E}$ states in the main text can be used as a reliable measure of their diabatic population.

\section{Discussion on the absence of the $\tilde{\mathrm{A}}$-state signal}

Identifying or excluding the presence of the $\tilde{\mathrm{A}}$-state in the experimental results is significantly easier if one has an estimate of the timescale of this state's dynamics. To obtain this we have performed quantum dynamical simulations of the state's vibrational dynamics by solving the time-dependent Schr{\"o}dinger equation (TDSE)
\begin{eqnarray} \label{eq.1}
   i \hbar \frac{\partial}{\partial t} \chi(R,t) & = &  \hat{H} \chi(R,t), \\
   \text{with}~\hat{H} & = & -\frac{1}{2 m_R} \frac{\partial^2}{\partial R^2} + \hat{V}_{X}(R),
\end{eqnarray}
where $m_r$ is the reduced mass along the dissociation coordinate $R$ and $\hat{V}_{X}(R)$ is the potential energy operator. The numerical propagation on the adiabatic PEC is performed by integration of the TDSE according to
\begin{equation}
 \chi(t + dt) = e^{(-i\hat{H} dt)} \chi(t)  = \hat{U} \chi(t).
\end{equation}
The evolution operator $\hat{U}$ is expanded in a Chebyshev series\cite{Tal-Ezer:1984}. The PEC used is represented on a one-dimensional spatial grid with 256 grid points and obtained by interpolating the results of the relaxed scan of cationic trifluoroiodomethane (\ref{fig:cf3i_relaxed_scan}). Due to the anti-bonding character of the $\tilde{\mathrm{A}}$ state, this is entirely repulsive. The limits of the grid used are \qty{1.6}{\angstrom} and \qty{3.4}{\angstrom}, respectively, and a Butterworth filter~\cite{Butterworth:1930} is employed, which absorbs the parts of the wavepacket that reach the dissociation area. Loss of population in the simulations can then be directly associated with dissociation. The filter is of "left-pass" type (absorbing all parts on the right side of the grid) and placed at \qty{3.3}{\angstrom} with an order of 100. For the simulation a time step of \qty{2}{\au} is used, and the simulation time is \SI{250}{\femto\second}. The propagation is initialized as the first eigenfunction of the neutral ground state potential to the $\tilde{A}$ state  potential, i.e. assuming delta-pulse ionization. The quantum dynamical simulations are conducted with QDng a program developed in-house\cite{kowalewski:2024}.\\

\begin{figure}[htb]
    \centering
    \includegraphics[width=\textwidth]{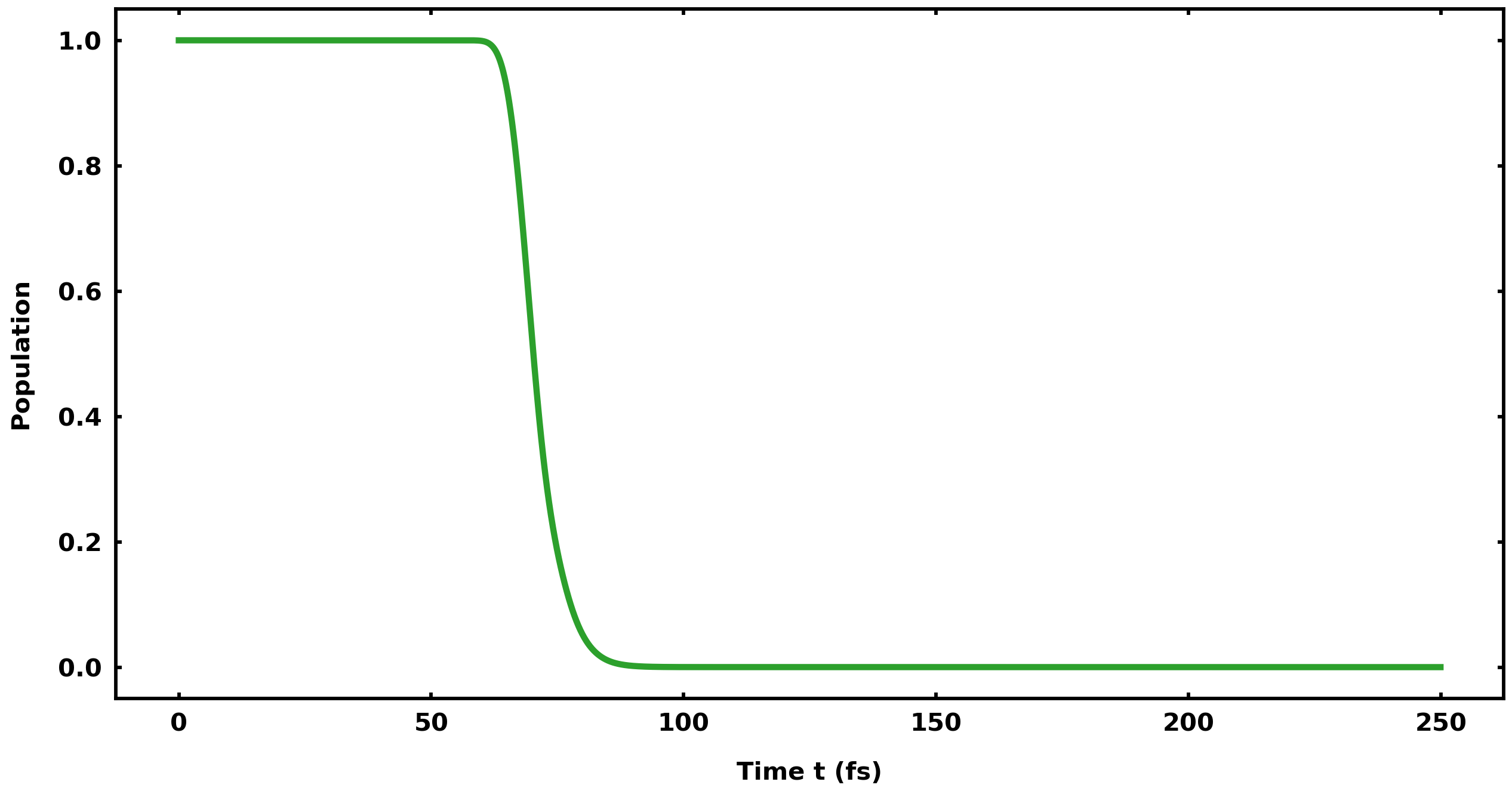}
    \caption{\label{fig:cf3i_population}Population of the $\tilde{A}$ state of the \ce{CF3I+} cation over the \qty{250}{\femto\second} of simulation time.}
\end{figure}

Figure~\ref{fig:cf3i_population} shows the change in population of the $\tilde{A}$ state over the \qty{250}{\femto\second} of simulation time. Within about \qty{70}{\femto\second} the $\tilde{A}$ state completely dissociates.

\begin{figure} [h]
    \centering
    \includegraphics[width=\textwidth]{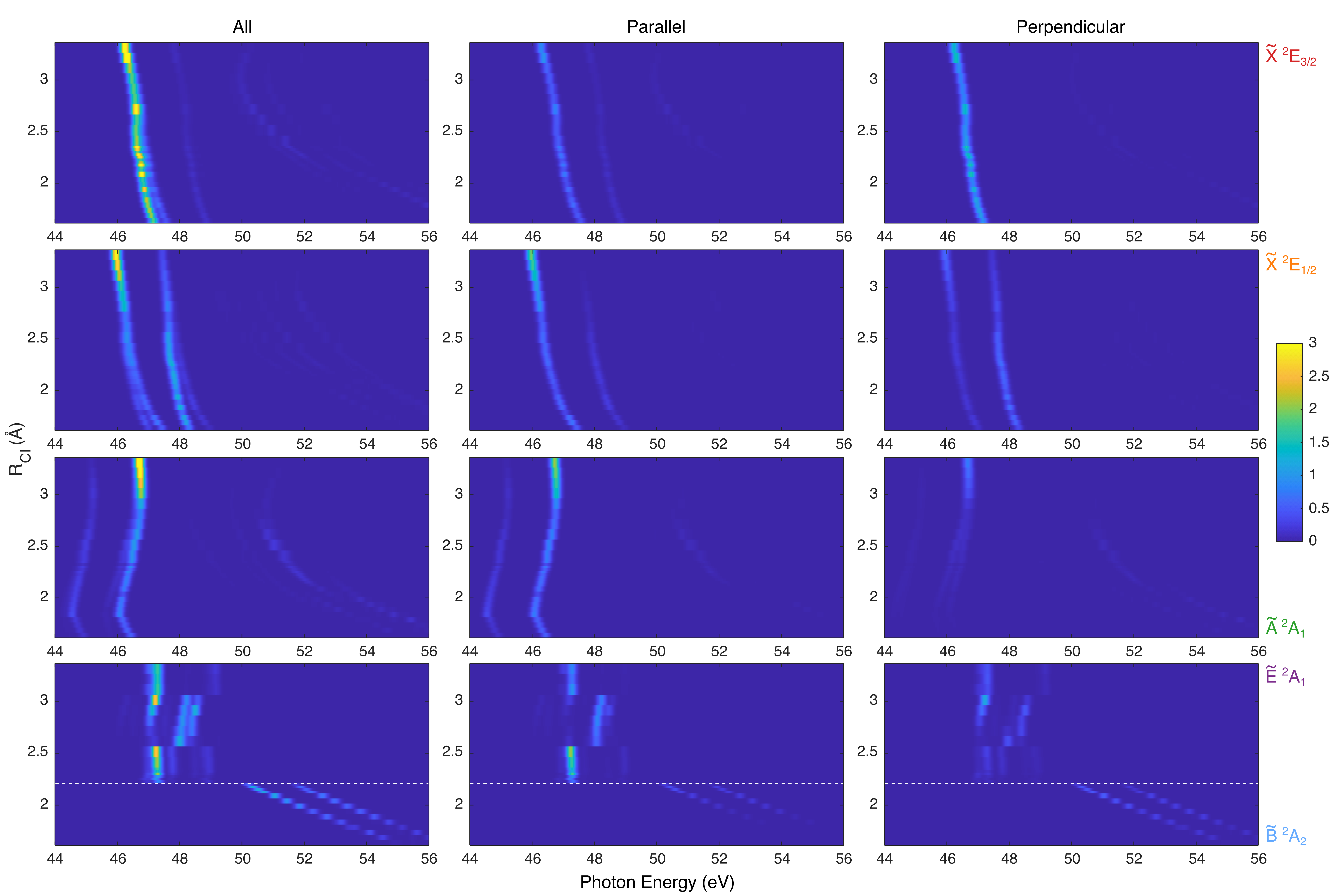}
    \caption{\label{fig:XAS_all}\textbf{State- and orientation-resolved XAS.} A common colorscale is used for all panels. The panels are labeled according to their orientation on the top, and according to their state on the right.}
\end{figure}

Turning now to the experimental results, we find no transient signals which exhibit both such a timescale and matching the calculated XAS. Although the absorption feature at 46.1~eV shows spectral overlap with the largest peak of the theoretical $\tilde{\mathrm{A}}$-state spectrum close to the FC point (see Figure~2F), its temporal evolution matches that of the $\tilde{\mathrm{X}}_{1/2}$ state far better --- it exhibits a very slight red shift on a 100~fs timescale. The $\tilde{\mathrm{A}}$-state should instead exhibit a rapid blue-shift, on the previously-identified 70~fs timescale, converging to 46.70~eV --- the absorption of the iodine $^2\mathrm{P}_{1/2}$ state (see Fig.~\ref{fig:XAS_all}). For this reason it is not assigned to the $\tilde{\mathrm{A}}$ state.

We suggest two explanations for the absence of significant $\tilde{\mathrm{A}}$-state absorption in our experimental results. The first reason combines the dependence of the strong-field ionization probability on the relative orientation of a molecule with the fact that the $\tilde{\mathrm{A}}$-state's XAS is dominated by parallel transitions. Alignment through orientation-sensitive ionization is a phenomenon that has been identified in ATAS before, for example in bromomethane~\cite{Timmers2019}. If the $\tilde{\mathrm{A}}$-state were to be created perpendicular to the pump and probe beam's polarisation axis, the XUV probe would be almost blind to the presence of the population, explaining why the $\tilde{\mathrm{A}}$-state signal is not observed.

A second explanation stems from the fact that the strong-field ionisation process is very difficult to predict accurately, especially in the presence of spin-orbit coupling, multiple ionisation channels and when the parent population is significantly depleted (as is the case in our experiment). Under such conditions, the probability of creating the $\tilde{\mathrm{A}}$-state in the first place may be significantly suppressed, leading to a lack of a measurable signal.

The correct explanation can be uncovered with further ATAS measurements which rotate the pump-polarization relative to that of the probe, or by introducing a third laser pulse to impulsively align the CF$_3$I prior to the ATAS experiment. Both of these efforts are currently underway but fall outside the scope of this work.

\clearpage

\section{Symmetry analysis of the CT reaction}

In this section we shall make use of symmetry arguments to draw conclusions about the number and symmetry of states and vibrational modes involved in the CT-reaction. This will allow us to infer the reaction path and explain the RASPT2 PECs and those of the three-state model.

First let us consider the vibrational problem. The point group of the neutral CF$_3$I molecule, as well as that of its cationic ground state is C$_{\rm 3v}$. The nine vibrational modes of the system consist of three doubly-degenerate E modes and three singly degenerate A$_1$ modes. CF$_3$I does not possess any A$_2$-symmetry modes. 

The RASPT2 calculations reveal the symmetry of the cationic electronic states involved in the CT reaction. We find the donor $\tilde{\mathrm{B}}$ state to be of $^2$A$_2$ symmetry and the acceptor $\tilde{\mathrm{E}}$ state to be of $^2$A$_1$ symmetry. The RASPT2 calculations also reveal the leading electronic character of the states (see Table~\ref{tab:electronic_character} and Figure~2B of the main text), which show that the transition between $\tilde{\mathrm{B}}$ and $\tilde{\mathrm{E}}$ requires the rearrangement of two electrons. As we are able to experimentally observe a non-zero population transfer delay, we are able to conclude that the CT-reaction passes through one (or several) intermediate diabatic state(s), collectively labeled $\tilde{\mathrm{I}}$. By splitting up the rearrangement into two single electron transitions we can identify the leading electronic character of the possible intermediate states, which we find to be of 
(lp$_3$)$^1$($\sigma_4$)$^2$(lp$_{1,2}$)$^3$($\sigma^*_5$)$^1$
and 
(lp$_3$)$^2$($\sigma_4$)$^2$(lp$_{1,2}$)$^3$($\sigma^*_5$)$^0$.
Both of these configurations can only yield states of $^2$E symmetry, informing us of the symmetry of the $\tilde{\mathrm{I}}$ state(s). 

Having identified the symmetry of all three diabatic states ($^2$A$_2$, $^2$E, $^2$A$_1$), we can infer the symmetry of the vibrational modes that might couple them. As the three states are of different symmetries, their direct coupling is not possible by any totally symmetric (A$_1$) vibrational mode, such as the C-I stretching mode, at C$_{3v}$ geometries. The reaction must therefore occur with participation of the remaining E vibrational modes.

While the E vibrational modes are able to couple $\tilde{\mathrm{B}}$ to $\tilde{\mathrm{I}}$ and $\tilde{\mathrm{I}}$ to $\tilde{\mathrm{E}}$, linear vibronic coupling between $\tilde{\mathrm{B}}$ and $\tilde{\mathrm{E}}$ mediated by an E-symmetry mode is forbidden because \cite{domcke04a,woerner09b}
\begin{equation*}
\mathrm{A}_1 \otimes \mathrm{A}_2 = \mathrm{A}_2 \neq \mathrm{E}
\end{equation*}
 and it is also forbidden by quadratic vibronic coupling by an E-symmetry mode because the symmetric part of $\mathrm{E}\otimes \mathrm{E}$ does not contain the direct product (A$_2$) of the irreducible representations of the $\tilde{\mathrm{B}}$ and $\tilde{\mathrm{E}}$ states:
\begin{equation*}
    [\mathrm{E}\otimes \mathrm{E}] = \mathrm{A}_1 \oplus \mathrm{E} \not\,\owns \mathrm{A}_2. 
\end{equation*}

The lowest order of vibronic coupling that allows for direct coupling of $\tilde{\mathrm{B}}$ and $\tilde{\mathrm{E}}$ is a bilinear coupling, requiring the simultaneous involvement of two E vibrational modes. The high-order nature of this interaction lends further support to the assumption that the direct coupling element ${V}_{13}$ in our effective one-dimensional three-state model is negligibly small. 

Compared to bilinear direct coupling, the indirect three-state coupling presented in our work is of lower order (requiring only a linear coupling in the E mode) and is the only mechanism that can reproduce a population transfer delay. From our symmetry analysis we can conclude that the one-dimensional reaction coordinate $x$ in Figure~5 represents a reaction path that runs along the A$_1$ C--I stretching mode but includes a finite displacement along one or several E-symmetry vibrational mode(s) such that it misses the CoIn. 

Since the RASPT2 caluclations are performed along a coordinate that preserves the C$_{\rm 3v}$ geometry of the ionic ground state, the $\tilde{\mathrm{B}}$/$\tilde{\mathrm{E}}$ adiabats cannot contain the $\tilde{\mathrm{I}}$ intermediate state character (and exhibit a very sudden change of character at the CoIn). The 3-model system adiabats, on the other hand, are not bound to the C$_{\rm 3v}$ point group or required by symmetry to exhibit a discontinuity in their electronic character. Coupling to the intermediate state in the three-state model is allowed; as a result they exhibit a more gradual change of electronic character and considerable NAC.